\documentclass[]{pasj01_revised}

\usepackage{graphicx,color}
\usepackage[OT2,T1,OT1]{fontenc}
\usepackage{lmodern}
\usepackage{natbib}

\begin{document} 

\title{Formation, Evolution, and Revolution of Galaxies by SKA: 
Activities of SKA-Japan Galaxy Evolution Sub-SWG}

\author{Tsutomu T.\ \textsc{Takeuchi}\altaffilmark{1}, 
Kana \textsc{Morokuma-Matsui},\altaffilmark{2}, 
Daisuke \textsc{Iono}\altaffilmark{2,3},
Hiroyuki \textsc{Hirashita}\altaffilmark{4},
Wei Leong \textsc{Tee}\altaffilmark{4,5},
Wei-Hao \textsc{Wang}\altaffilmark{4},
Rieko \textsc{Momose}\altaffilmark{2,6,7}, 
on behalf of the SKA-Japan Galaxy Evolution sub-Science Working Group
}

\altaffiltext{1}{Division of Particle and Astrophysical Science, Nagoya University, 
Furo-cho, Chikusa-ku, Nagoya 464--8602, Japan.}
\email{takeuchi.tsutomu@g.mbox.nagoya-u.ac.jp}

\altaffiltext{2}{National Astronomical Observatory of Japan, Osawa, Mitaka, Tokyo, Japan}
\email{kana.matsui@nao.ac.jp}

\altaffiltext{3}{SOKENDAI (The Graduate University for Advanced Studies) }
\email{d.iono@nao.ac.jp}

\altaffiltext{4}{Institute of Astronomy and Astrophysics, Academia Sinica,
P.O. Box 23-141, Taipei 10617, Taiwan}
\email{hirashita@asiaa.sinica.edu.tw}

\altaffiltext{5}{Department of Physics, National Taiwan University, Taipei 10617, Taiwan}

\altaffiltext{6}{National Tsing Hua University, Hsinchu, 30013, Taiwan}

\altaffiltext{7}{The Institute for Cosmic Ray Research, the University of Tokyo, Japan}
\email{momose.rieko@nao.ac.jp}





\maketitle

\begin{figure}
\vspace{-238mm}
\includegraphics[width=25mm]{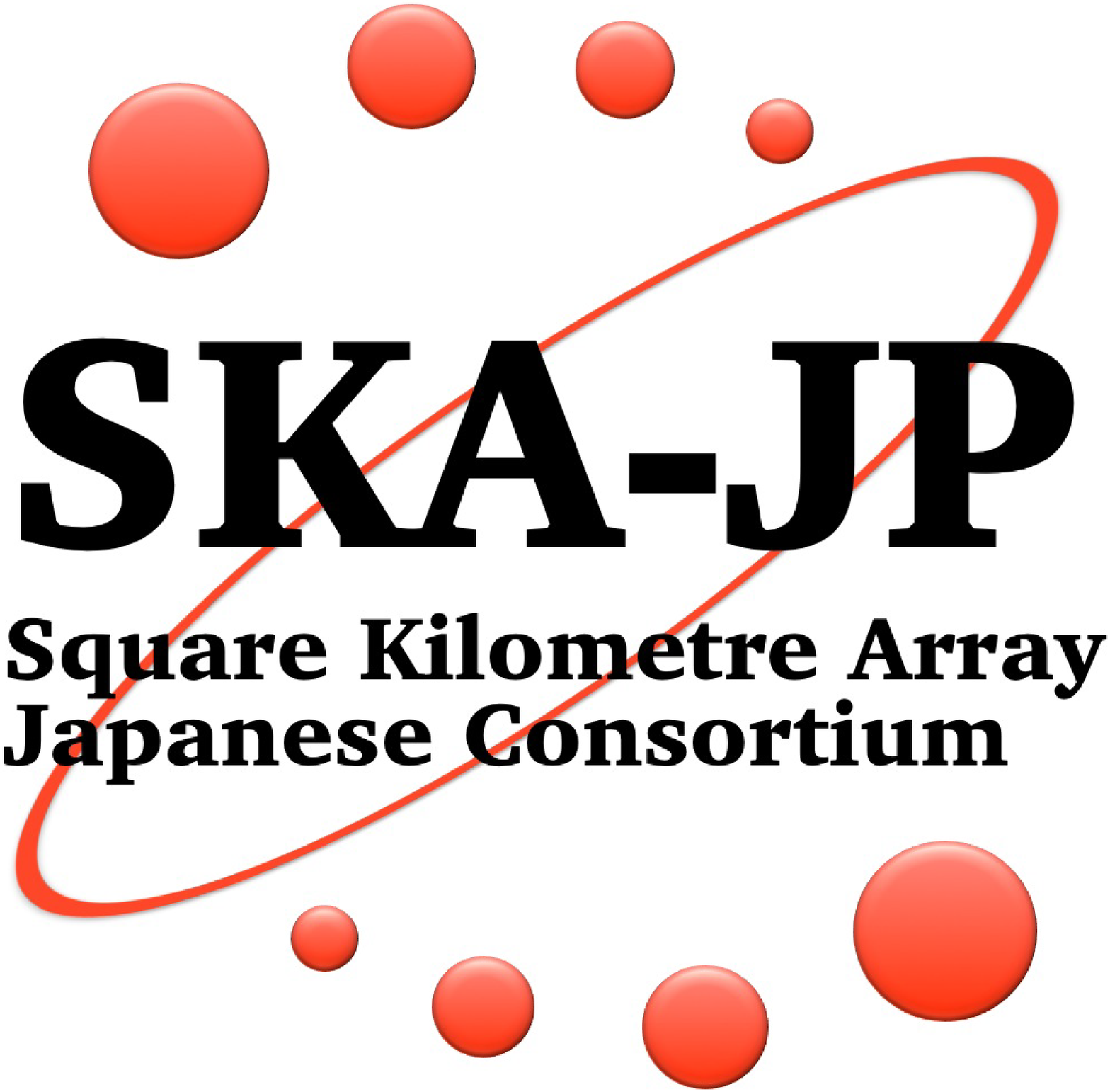}
\vspace{177mm}
\end{figure}

\begin{abstract}
Formation and evolution of galaxies have been a central driving
force in the studies of galaxies and cosmology. Recent studies
provided a global picture of cosmic star formation history.
However, what drives the evolution of star formation activities
in galaxies has long been a matter of debate. The key factor of
the star formation is the transition of hydrogen from atomic to
molecular state, since the star formation is associated with
the molecular phase. This transition is also strongly coupled
with chemical evolution, because dust grains, i.e., tiny solid
particles of heavy elements, play a critical role in molecular
formation. 
Therefore, a comprehensive understanding of neutral--molecular 
gas transition, star formation and chemical enrichment
is necessary to clarify the galaxy formation and evolution.
Here we present the activity of SKA-JP galaxy evolution sub-science working group 
(subSWG)
Our activity is focused on three epochs: $z \sim 0$, 1, and $z > 3$. 
At $z \sim 0$, we try to construct a unified picture
of atomic and molecular hydrogen through nearby galaxies in
terms of metallicity and other various ISM properties. Up to
intermediate redshifts $z \sim 1$, we explore scaling relations
including gas and star formation properties, like the main
sequence and the Kennicutt--Schmidt law of star forming galaxies. 
{}To connect the global studies with spatially-resolved investigations,
such relations will be plausibly a viable way. For high redshift
objects, the absorption lines of HI 21-cm line will be a very
promising observable to explore the properties of gas in galaxies.
By these studies, we will surely witness a real revolution in
the studies of galaxies by SKA. 
\end{abstract}

\section{Introduction}

Galaxies evolve. 
This fact has been a driving force of the studies on galaxies from late 70's when
the galactic physics started kicking in, until today.
However, how galaxeis formed and evolved still remains far from full 
understanding. 
The evolution of galaxies is a change of their various complicated physical 
properties with time. 
{}To be precise, there are two different aspects in galaxy evolution: 
the individual evolution of a certain galaxy, say, a ``personal''
history of a galaxy, and the statistical evolution of galaxies at a certain cosmic time, 
or a ``social'' history of galaxies. 
We use one of these different aspects depending on the issue we are concerned.
{}From a cosmological point of view, the dark matter interacts gravitationally to form
dark halos, and galaxies form and evolve in the halos. 
In this case, the dynamical properties are most important.
However, the most enthusiastically studied subject of galaxy evolution is the
evolution of star formation (SF) activities: 
stars form in a first gas clump, and first generation stars die and expel the first
heavy elements in the Universe, and the formation of the next generation stars 
is accelerated, and so on. 
The chemical composition of the interstellar medium (ISM) in galaxies changes
with time (chemical evolution: \citealt{tinsley1980}). 
In addition to the snapshot of the SF, the accumulated stellar mass became
of interest recently. 
Not only the evolution of galaxies, but also the mass accretion in the form of  
dwarf galaxies and remnant gas infall from filamentary structures in the Large-Scale Structure in
the Universe change the structure of galaxies.
Galaxy merging drastically changes the morphology of galaxies, which is also one of the
important aspects of galaxy evolution. 
There often exists a massive black hole at the center of a galaxy. 
The activity of an active galactic nucleus (AGN) evolves with the change of 
the accretion rate to the black hole. 
A mysterious correlation between the masses of the spheroid component and 
central black hole in galaxies \citep{magorrian1998}. 
This suggests a hidden connection of the evolution of black holes and galaxies,
known as ``the co-evolution''. 

During the recent decade, multiwavelength deep galaxy surveys brought us 
a dramatic view of very high-$z$ Universe. 
They seem to show that the properties of galaxies are mainly determined by
their stellar mass and environment, among others.
According to the standard cold dark matter (CDM) scenario, the structure
formation proceeds hierarchically, i.e., small halos form tend to form early and
they merge each other to form larger, heavier objects.
Before galaxy formation, baryons are mostly in gas phase. 
Then, they fall onto the gravitational potential of halos and form dense molecular
clouds. 
This is the first onset of star formation.  
Dying stars inject metals and dust into the ISM, which accelerate the next
star formation. 
At the same time, they also provide thermal and kinetic energy to the ISM 
which hamper the cooling of gas. 
Thus, the star formation is self-regulating, very complicated process determined
by the interplay of DM and baryons. 
Hence, in order to understand the SF, we must observe the neutral gas in 
galaxies in various environment as a function of the cosmic time, as many as possible. 

However, previous studies on galaxy evolution were dominantly focused on the
snapshot of the SF activities in galaxies. 
The star formation rate (SFR) is observationally estimated by various observables
related to massive stars [recombination lines of H{\sc ii} regions, 
non-ionizing ultraviolet (UV) luminosity, dust emission heated by UV from massive
stars, polycyclic aromatic hydrocarbon (PAH) band emission, radio continuum, etc.], 
but all of them can only give the information of the SFR, and nothing about {\it the
transition from gas to stars}, the fundamental key process of the SF. 
In radio astronomy, in contrast, traditionally the molecular emission lines have
been observed to obtain the information of the material to form stars. 
Combining the SFR and molecular gas mass, we can discuss the gas--star
transition.
But still the molecular lines can tell us only about gas in molecular form, 
and if we want to go further to have a global view of the SF, we must observe
the neutral gas. 
Up to now, because of the limitation of the current instruments, the observation of 
neutral gas was limited to nearby galaxies, and the properties of neutral gas 
in galaxies have been rarely discussed in the context of galaxy evolution.
The advent of the square kilometre array (SKA) will bring the ultimate, revolutionary 
breakthrough to the studies on galaxy evolution. 

This article is organized as follows: in Section~\ref{sec:HI}, we overview the role of 
atomic hydrogen (H{\sc i}) in the context of galaxy evolution. 
We also summarize the SKA-related researches on galaxy evolution
in this section. 
In Section~\ref{sec:SKA_JP}, we introduce the activity of the SKA-JP 
galaxy evolution sub-science WG. 
Section~\ref{sec:summary} is devoted to summary.

\section{Current Status of the Studies on Galaxy Evolution --- Atomic Hydrogen in Galaxies ---}
\label{sec:HI}

\subsection{Atomic and molecular hydrogen}

Molecular gas is the key ingredient for star formation in galaxies.  
Roughly 10\% of the mass of the interstellar medium is composed of molecular hydrogen (H$_2$) in the Milky Way galaxy \citep{kk09,ns15}.
This subsection describes the processes responsible for the formation and destruction of 
molecular hydrogen, and the fractional abundance of atomic and molecular hydrogen.  While the principle target line for SKA
will be atomic hydrogen, both atomic and molecular lines are intimately connected as seen in the following section.  
Deep understanding of both phases of the ISM is critical. 

\subsubsection{Formation of molecular hydrogen}

The basic formation path of molecular hydrogen in galaxies is the coupled reaction between two hydrogen atoms, which
occurs using electrons and protons as a catalyst. This reaction is accompanied by a significant amount of exothermic energy 
release.  Unless this exothermic energy is deposited into some form of energy, the reaction will reverse and molecules will not form.
Therefore, a third atomic hydrogen that will collisionaly remove the excess energy is necessary in order to form the hydrogen molecule.
This three hydrogen-atom reaction is efficient in very dense (n$_{\rm H} > 10^8$~cm$^{-3}$) environments.  
In the more diffuse galactic ISM (n$_{\rm H} \sim 10^5$~cm$^{-3}$), 
molecular hydrogen forms using interstellar dust as the catalyst \citep{gs63,hs71}.   
It is believed that the excess energy generated from the bonding of the two 
hydrogen atoms is absorbed by the random motion of the interstellar dust.

\subsubsection{Dissociation of molecular hydrogen}

Molecular hydrogen can be dissociated and it can result in two atomic hydrogen.  There 
are three mechanism for this dissociation process.  First is the photo-dissociation due to 
UV photons (Lyman-Werner photons) that are emitted from O or B type stars in the 
range of 11.2 - 13.6 eV.  
The second process is the photo-dissociation due to cosmic rays that have 1--100~MeV of energy.
The third process is the collisional dissociation in high temperature high density regions.  
However, it is thought that 
the collisional dissociation process itself has a minor contribution to the dissociation of 
molecular hydrogen.  This is because molecular hydrogen formation occurs more efficiently
than dissociation, due to the efficient depositing of thermal energy generated by molecular hydrogen
dissociation and cooling by metals in the ISM. 

\subsubsection{Atomic and molecular hydrogen in galaxies}

The fraction between atomic and molecular gas ($f_{\rm mol} = \rho_{\rm H_2}/\rho_{\rm total}, \; \rho_{\rm total}=\rho_{\rm H_2}+\rho_{\rm HI}$)
is determined by the balance between
formation and dissociation of molecular hydrogen.  \citet{elmegreen93} proposed a 1-D model in which
the $f_{\rm mol}$ is described by the pressure, UV radiation field, and metalicity in the molecular cloud.
\citet{krumholz08} expanded \citet{elmegreen93}'s model using a spherically symmetric 
multi-dementional model, and added the formation of molecular hydrogen from interstellar dust 
as a new parameter.  These models reproduce the observations of nearby galaxies and our own 
galaxy \citep{sofue95, honma95, br04, leroy08}.

Observational check of $f_{\rm mol}$ is mainly performed using the $^{12}$CO~(J=1--0) line (2.6 mm) emission
and the H\emissiontype{I} (21 cm) line emission.  The $^{12}$CO~(J=1--0) line allows us to derive the
line strength and the molecular hydrogen surface density via the conversion factor 
($X_{\rm CO}$ [cm$^{-2}$(K km s$^{-1}$)$^{-1}$]), and the H\emissiontype{I} line allows us to obtain the 
surface density of atomic hydrogen.  The $f_{\rm mol}$ is derived from these surface densities.
The $f_{\rm mol}$ is 25--30~\% in late type galaxies \citep{boselli02}.  
Furthermore, the 
radial distribution of the $f_{\rm mol}$ decreases as a function of radius \cite[e.g.,][]{bb12, tanaka14}.

We now describe the density that is required for the transition from atomic to molecular gas.  
According to analytical models, shielding of atomic hydrogen from photodissociation 
occurs effectively at $\Sigma_{\rm HI} \sim 10$~M$_{\odot}$~pc$^{-2}$ ($N_{\rm H} \sim 10^{21}$ cm$^{-2}$)
in a spherical molecular cloud with solar metalicity \citep{krumholz09}.  In galaxy evolution 
simulations including molecular hydrogen formation and dissociation, almost all of the 
atomic hydrogen becomes molecular at $N_{\rm H} \sim 10^{21}$~cm$^{-2}$ in galaxies 
with similar metallicities \cite[e.g.,][]{pelupessy06, gnedin09}.  Observations of
late type galaxies with solar metallicities show similar results to these theoretical predictions
\citep[e.g.,][]{wb02, bigiel08}.

The required column density for molecular gas formation depends on the metallicity.
\citet{gnedin09} suggest that the required column density is higher for low metallicity
environments.  In their simulations, the dust particles required for shielding from photo-dissociation
decreases at low metallicity environments.  Therefore, the column density required for shielding 
becomes higher.  Similar results are obtained from other theoretical models and observations 
of nearby galaxies \citep[e.g.,][]{krumholz09, mk10, fumagalli10, wong13, richings14}, 
and thus the metallicity appears to be the key parameter
that controls the process that converts from atomic to molecular hydrogen.

\subsection{Atomic hydrogen in spiral galaxies}
\label{sec:spiral}

Observations of atomic hydrogen gas (H\emissiontype{I} gas) in spiral galaxies 
goes back to the 1950's.  Ever since, a large number of galaxies have been observed from 
very nearby ($\sim 50$~kpc) to distances larger than 100~Mpc ($z > 0.024$), using 
single dish and interferometers.  A summary of the physical understanding of atomic gas 
in galaxies obtained through H\emissiontype{I} observations in spiral galaxies is presented in this subsection. 

\subsubsection{Spatial distribution}

There are three major characteristics in the distribution of H\emissiontype{I} in spiral galaxies.  First is the 
decrease in the galactic center, where the density of the ISM is high.  Therefore, most of the 
hydrogen is in molecular form, and the abundance of H\emissiontype{I} gas is very low.

The second characteristic is that the H\emissiontype{I} gas is more extended than the optical disk 
\citep[e.g.,][]{sancisi83, leroy08}).  
As such, H\emissiontype{I} gas is often 2-3 times 
more extended than disk stars in many of the spiral galaxies \citep[e.g.,][]{sancisi83}.
The H\emissiontype{I} gas seen in the outskirts of spiral galaxies are called circum-galactic medium (CGM).
Theoretically, the CGM is thought to be related to large scale gas accretion or gas outflow 
\cite[e.g.,][]{mori02,mori06,dekel09a,dekel09b,scannapieco05}, 
and they are intimately connected to the mass accretion process and 
star formation history in galaxies.  It is difficult to directly detect the 
gas in the CGM, mainly because the column density of H\emissiontype{I} gas in the CGM is lower than the disk
gas ($N_{\rm HI} < 10^{19}$~cm$^{-2}$). However, the advancement in recent receivers have allowed deep and 
sensitive H\emissiontype{I} observations, and the diffuse H\emissiontype{I} gas ($N_{\rm HI} 
\sim 10^{18}$~cm$^{-2}$) in the CGM 
has been detected \citep{wolfe13,pisano14}.  Possibility of gas inflow onto spiral galaxies have been 
proposed based on these new data.

The third characteristics is that the surface density ($\Sigma_{\rm HI}$ [M$_{\odot}$ pc$^{-2}$]) of H\emissiontype{I} in the
exterior of spiral galaxies are fairly constant and decreases at the exterior.  The $\Sigma_{\rm HI}$ is low at the center, and increases 
as a function of radius, peaking at some radius ($\Sigma^{\rm max}_{\rm HI}$).  The $\Sigma_{\rm HI}$ 
beyond this peak radius is fairly flat and constant \citep[e.g.,][]{sofue01, leroy08, bb12}. 
 $\Sigma^{\rm max}_{\rm HI}$ only goes as high as $\sim 10$~M$_{\odot}$~pc$^{-2}$ because atomic hydrogen
 becomes molecular when $\Sigma^{\rm max}_{\rm HI} > 10$~M$_{\odot}$~pc$^{-2}$. 

\subsubsection{H\emissiontype{I} gas mass}

The observed H\emissiontype{I} gas mass in spiral galaxies is M$_{\rm HI} \sim 10^{8-10}$~M$_{\odot}$
\citep{walter08}.  In particular, sources with M$_{\rm HI} \sim 10^{10}$~M$_{\odot}$ are often 
colliding galaxies, and therefore galaxy interaction can bring significant amount of material into the galaxy.

It is known that the total H\emissiontype{I} mass and the disk size in spiral galaxies show a correlation. As described in 
\S4.3.2, $\Sigma_{\rm HI}$ in the exteriors of spiral galaxies is almost constant.  If we assume that 
$\Sigma_{\rm HI}$ is constant across the galaxy disk, then the total H\emissiontype{I} mass of a spiral galaxy is roughly 
proportional to the area of the galaxy.  In fact, it is suggested from observations 
that the total H\emissiontype{I} mass correlates with the radius of the stellar disk \cite[e.g.,][]{bb12}).\\

\subsubsection{Kinematical information}

One can determine the mass distribution of the galaxy from the kinematical information provided by HI observations.  
While the rotation velocity varies from 150 - 300~km~s$^{-1}$ on average, 
the basic characteristic is that the rotation increases as a function of radius until some radius, beyond which
the velocity is nearly constant ($v_{\rm max}$) 
\citep[e.g.,][]{rubin80, rubin85, bosma81, sancisi1987a, sancisi1987b}, due to dark matter.

\subsubsection{Tully--Fisher relation}

The rotation velocity and the brightness of spiral galaxies are correlated, in what is called the 
Tully-Fisher relation \citep[][TF relation:]{tf77}.  
It is known that the scatter in the TF relation is larger at the low mass end.  This is due to the fact that
the fraction of gas mass becomes non-negligible for low mass spiral galaxies.  The correlation becomes
tighter when the total baryon mass ($M_{\rm baryon} = M_{\rm star} + M_{\rm gas}$) is used instead of the 
stellar mass \citep[e.g.,][]{bdj01}.  This is called the baryon-Tully-Fisher relation (BTF relation).
Observational confirmation of the BTF relation is currently ongoing, using the recent gas survey observations
of spiral galaxies \cite[e.g.,][]{mcgaugh12}.

The TF relation is used to measure the distance to spiral galaxies.  The true brightness of the galaxy 
can be estimated from the TF relation and the H\emissiontype{I} rotation curve.  On the other hand, the apparent brightness 
of the galaxy depends on the distance.  Therefore, the difference between the apparent and the true
brightnesses allows us to determine the distance to the spiral galaxy.   Up to now, the H\emissiontype{I} velocity and the
stellar spectroscopy were used to estimate the distances.  
More recently, the BTF relation has been used 
to get an accurate measure of the distances \citep[e.g.,][]{bdj01}.  
However, significant uncertainties
are present in the derivation of the stellar and gas masses, and therefore uncertainties in the derived 
distances \citep[e.g.,][]{zaritsky14}.

\subsection{Atomic hydrogen in early-type galaxies}

\label{sec:etg}

Early type galaxies (ETGs, elliptical and lenticular galaxies) have been considered to be characterized by higher bulge-to-disk ratio and older stellar population compared to late type galaxies (LTGs, spiral galaxies).
Recent observations revealed the variety of ETGs in morphology, e.g. existence of ETGs with bulge-to-disk ratio comparable to LTGs \citep{1951ApJ...113..413S,1970ApJ...160..831S,1976ApJ...206..883V}, as well as in star formation activity \citep[e.g.,][]{2010MNRAS.408...97K}.
In this section, we take a look back to simple histories of H\emissiontype{I} observations targeting ETGs and introduce some science outputs from ATLAS$^{\rm 3D}$, one of the largest surveys toward ETGs.

\subsubsection{Single-dish studies}

ETGs have been observed in H\emissiontype{I} since 1960 era.
Although interferometric observation can achieve higher angular resolution than single-dish observation, single-dish H\emissiontype{I} observations enable us to measure total amount of atomic gas and line-width of galactic H\emissiontype{I} spectra without so-called ``missing-flux'' problem (interferometer does not have a sensitivity on extended structure).
In 1960s, H\emissiontype{I} observations of ETGs revealed that the ratios of H\emissiontype{I} mass to $B$-band luminosity of ETGs are lower than those of LTGs \citep[e.g.,][]{1969A&A.....3..281G}.
Based on these observations, H\emissiontype{I} gas masses of 15~\% of $\sim150$ ellipticals and 25~\% of $\sim300$ lenticulars were successfully measured with a typical detection mass limit of a few times $10^8$ M$_\odot$ \citep{1985AJ.....90..454K,1986AJ.....91...23W}.

In 2000s and beyond, H\emissiontype{I} blind surveys, such as HIPASS \citep{2001MNRAS.322..486B} and ALFALFA \citep{2005AJ....130.2598G} have revealed a strong environmental dependence of H\emissiontype{I} detection rate:
a few~\% in the galaxy cluster environment whereas 40~\% in the field environment \citep{2007A&A...474..851D,2009A&A...498..407G}.
These results are consistent with the idea of \citet{1983AJ.....88..881G} that H\emissiontype{I} gas is easily stripped from galaxies in cluster environment as well as the other recent observations showing that star-forming ETGs reside in low galaxy density region \citep{2010MNRAS.404.1775T}.


\subsubsection{Interferometric studies}

In 1980s, detailed spatial distribution of H\emissiontype{I} gas in ETGs have been investigated using interferometric data.
Despite the appearance in optical images, ETGs show a wide variety of H\emissiontype{I} spatial distribution, such as large disk extending to tens of kpc, low-density disk, ring structure, and unsettled gas distributions indicating recent gas accretion, gas stripping, galaxy interaction and merging \citep[][and reference therein]{1997ASPC..116..310V,2001ASPC..240..657H}.
These study revealed that H\emissiontype{I} gas is one of the ideal tracers of recent mass assembly history of ETGs.

With Westerbork Synthesis Radio Telescope (WSRT), \citet{2006MNRAS.371..157M} and \citet{2010MNRAS.409..500O} conducted deep H\emissiontype{I} imaging observations of 33 ETGs selected from SAURON project\footnote{
SAURON (Spectroscopic Areal Unit for Research on Optical Nebulae) is a name of the panoramic integral field spectrograph as well as a name of the scientific project whose main goal is to understand the formation and evolution of elliptical and lenticular galaxies and of spiral bulges from 3D observations \citep{2001MNRAS.326...23B}.
},
which is deep enough to detect H\emissiontype{I} mass down to $\sim(2-3)\times10^6$ M$_\odot$.
They confirmed environmental dependence of H\emissiontype{I} detection rate, where H\emissiontype{I} emissions are detected from 10~\% and $2/3$ of all ETGs inside and outside the Virgo cluster, respectively.
They revealed that the faintest signature of H\emissiontype{I} accretion is presented in most H\emissiontype{I}-detected ETGs.
In addition, they find that all ETGs with a settled H\emissiontype{I} distribution host ionized and molecular gas, suggesting recent star formation.

\subsubsection{H\emissiontype{I} survey towards ATLAS$^{\rm 3D}$ galaxies}

\citet{2012MNRAS.422.1835S} observed 170 ETGs in H\emissiontype{I} with WSRT as a part of ATLAS$^{\rm 3D}$ project.
The ATLAS$^{\rm 3D}$ project combines multi-wavelength surveys specialized in ETGs \citep{2011MNRAS.416.1680C}.
This project targets a complete sample of 260~early-type galaxies with $K$-band absolute magnitude of brighter than $-21.5$ within the local Universe (42~Mpc).
The achieved H\emissiontype{I} mass sensitivity is down to $(5-50)\times10^6$ M$_\odot$ and H\emissiontype{I} emissions were detected from 53 ETGs.
Based on this data, they revealed the properties of H\emissiontype{I} gas in ETGs as follows:
\begin{itemize}

\item A wide variety of H\emissiontype{I} morphology:
H\emissiontype{I}-detected ETGs shows disk and ring structures ($\sim64$~\%), unsettled structure including a tail structure which is indicative of tidal effect or gas accretion ($\sim26$~\%), and clumpy structure distributed over the entire galaxy ($\sim9$~\%).
ETGs with extended H\emissiontype{I} disk ($>3.5\times R_{\rm e}$) have large amount of H\emissiontype{I} gas ($5\times 10^9$ M$_\odot$) and different kinematics to stellar component, whereas ETGs with small H\emissiontype{I} disk have small amount of H\emissiontype{I} gas ($<10^8$ M$_\odot$) and the same kinematics with stellar component.

\item Signature of star formation:
70~\% of ETGs with H\emissiontype{I} gas within $\sim1 R_{\rm e}$ region show star formation activity, and their central regions are dominated by molecular gas.

\item H\emissiontype{I} mass function:
The characteristic mass is $\sim2\times10^9$ M$_\odot$ and the slope at lower mass is $-0.7$, when the H\emissiontype{I} mass function is fitted with a Schechter function.

\item Comparison with LTGs:
H\emissiontype{I} mass of ETGs is mostly lower than that of LTGs but some ETGs have H\emissiontype{I} mass comparable to that of LTGs.
The high-H\emissiontype{I}-column-density component observed in bright stellar disk of LTGs is not seen in ETGs.

\item Environmental effect on the nature of ETGs:
The environmental effect on H\emissiontype{I} gas mass in ETGs was confirmed, where cluster ETGs tend to have lower detection rate of H\emissiontype{I} emission ($\sim10$~\%) compared to field ETGs ($\sim40$~\%).
The H\emissiontype{I} gas mass and $M_{\rm HI}/L_K$ decrease as the number density of galaxies increases.

In addition, H\emissiontype{I} morphology correlate with the number density of galaxies, where ETGs with larger H\emissiontype{I} disk and ring structures tend to reside in lower galaxy density and ETGs with unsettled H\emissiontype{I} structure tend to reside in typical galaxy density of rich group.
This results indicate that processes working in galaxy-cluster scale are important for ETG evolution.

\end{itemize}

\subsection{Atomic hydrogen in dwarf galaxies}
\label{sec:dwarf}

Understanding star formation in dwarf galaxies is important step for understanding star formation in early Universe.
Dwarf galaxies are considered to reside in low-mass dark matter halos (DHs), and that the first star has been formed in such low mass DHs in $\Lambda$CDM model \citep{2006ApJ...652....6Y,2008Sci...321..669Y}.
In addition, revealing the nature of dwarf galaxies is also important for galaxy evolution since dwarf galaxies are considered to be a building block of larger galaxies.

Large galaxy surveys have been conducted mainly in optical and near infrared wavelengths, while these surveys tends to be biased to relatively bright and massive galaxies.
Recent H\emissiontype{I} galaxy surveys, on the other hand, have discovered a lot of gas-rich dwarf galaxies.
In this section, H\emissiontype{I} surveys specialized in dwarf galaxies are introduced.

\subsubsection{H\emissiontype{I} survey toward dwarf galaxies}

In the past 10 years, detailed kinematics and stellar population of dwarf galaxies with H\emissiontype{I} mass of $>10^8$ M$_\odot$ have been investigated.
ALFALFA \citep[Arecibo Legacy Fast ALFA Survey,][]{2005AJ....130.2598G}, which is a large and unbiased H\emissiontype{I} survey, discovered a few times 100 dwarf galaxies with H\emissiontype{I} mass $<10^8$ M$_\odot$, and provided us a statistically reliable H\emissiontype{I} mass function of galaxies for the first time \citep{2010ApJ...723.1359M}.
In the following paragraphs, four representative H\emissiontype{I} surveys towards dwarf galaxies with lower H\emissiontype{I} gas mass are introduced.

The FIGGS \citep[Faint Irregular Galaxies GMRT Survey,][]{2008MNRAS.386.1667B}, LITTLE THINGS \citep[Local Irregulars That Trace Luminosity Extremes, The H\emissiontype{I} Nearby Galaxy Survey,][]{2012AJ....144..134H}, and VLA-ANGST \citep[VLA A High-resolution H I Survey of Nearby Dwarf Galaxies,][]{2012AJ....144..123O} target dwarf galaxies with relatively small H\emissiontype{I} mass, $8.5\times10^7$, $2.7\times10^7$ and $2.3\times10^7$ M$_\odot$ as a median value, respectively.
21 dwarf galaxies with H\emissiontype{I} mass of $<10^7$ M$_\odot$ were also observed in those surveys (5, LITTLE THINGS; 7, FIGSS; 9, VLA-ANGST).
SHIELD \citep[Survey of H\emissiontype{I} in Extremely Low-mass Dwarfs,][]{2011ApJ...739L..22C} targets dwarf galaxies with $10^6-10^7$ M$_\odot$.
The scientific meanings of studying H\emissiontype{I} gas in dwarfs are as follows:
\begin{itemize}

\item Gas-rich dwarf galaxies:
H\emissiontype{I} surveys toward dwarf galaxies revealed the existence of gas-rich systems \citep{2008MNRAS.386.1667B,2012AJ....144..134H,2012AJ....144..123O}.
The existence of such gas-rich galaxies with very shallow potential wells poses interesting puzzles for the $\Lambda$CDM paradigm.
This is because that cold gas in shallow potential wells is expected to be easily striped by ram pressure \citep[e.g.,][]{2002MNRAS.334..673L,2003AJ....125.1926G}, be blown out by star formation activity \citep[e.g.,][]{1999ApJ...513..142M,2000MNRAS.313..291F}, and be vaporized by hot intergalactic medium \citep{2002MNRAS.333..156B}.
In addition, the UV radiation field inhibits gas accretion and cooling in low-mass halos \citep{1986MNRAS.218P..25R,1992MNRAS.255..346B,2002MNRAS.333..156B,2006MNRAS.371..401H}.
Studying such gas-rich dwarfs lead to the understanding of the gas deficiency processes of galaxies described above.

\item Star formation under the extreme environment:
Star formation is often induced by gas compression due to external perturbations such as galactic shock at the spiral arms in disk galaxies and galaxy mergers.
Dwarf galaxy is a suitable object to study star formation without these external perturbations.

\item Baryonic Tully-Fisher relation:
Dwarf galaxies do not follow the same Tully-Fisher relation of spiral galaxies.
However, once the baryon mass is used instead of galaxy luminosity, i.e., baryonic Tully-FIsher relation, they follow the same relation \citep{2000ApJ...533L..99M,2005ApJ...632..859M}, suggesting a close link between dark matter halo and baryonic components of galaxies.
Studying baryonic Tully-Fisher relation leads to the understanding of the relationship between evolutions of dark matter and baryonic components of galaxies.

\item Dark matter profile:
Dwarf galaxies are suitable targets to study the density profile of dark matter halos.
Cosmological simulations of hierarchical galaxy formation predict a ``universal'' cusped density core for the dark matter haloes of galaxies \citep[e.g.,][]{2004MNRAS.349.1039N}.
Although some observation support the theoretical prediction \citep[e.g.,][]{2001MNRAS.325.1017V,2005ApJ...634..227D}, other observations indicate a constant-density core for their dark matter haloes \citep[e.g.,][]{2003MNRAS.340..657D,2003MNRAS.340...12W}.
This is a so-called ``core-cusp problem''.
Dark matter profiles estimated by observations generally contain a significant error originated in the estimation of $M/L$.
However, it is possible to estimate dark matter profile of dwarf galaxy with a small contribution from stellar component.

\end{itemize}


\subsection{H\emissiontype{I} surveys covering various population of galaxies}

Investigating statistical properties of galaxies is also important for understanding galaxy evolution as well as detailed studies on individual galaxies.
In the previous subsections, we focused on the H\emissiontype{I} studies on spiral (sec.~\ref{sec:spiral}), early-type (sec.~\ref{sec:etg}) and dwarf galaxies (sec.~\ref{sec:dwarf}).
In this subsection, we first introduce one of the largest H\emissiontype{I} survey of local galaxies, GASS project, and then two pioneering H\emissiontype{I} surveys targeting galaxies in the ranges of $0.1<z<1$, BUDHIES and CHILES projects.




\subsubsection{GALEX Arecibo SDSS Survey (GASS)}

The bimodal distribution of galaxies in stellar mass versus $u-r$ color plot \citep[``galaxy bimodality'',][]{2012A&A...546A..22S} is one of the important scientific output from large galaxy surveys in optical wavelength, such as SDSS \citep[Sloan Digital Sky Survey,][]{2000AJ....120.1579Y}:
``blue clouds'' galaxies which are actively forming stars ($<3\times10^{10}$ M$_\odot$);
``red sequence'' galaxies which are not currently forming stars ($>3\times10^{10}$ M$_\odot$).
Galaxies distributed between the two sequences are referred to as ``green valley'' galaxies.

GALEX Arecibo SDSS Survey (GASS) is an H\emissiontype{I} survey specifically designed to measure H\emissiontype{I} mass of $\sim1000$ galaxies in the local Universe ($0.025 < z < 0.05$) with stellar masses $10^{10}<M_\star<10^{11.5}$ M$_\odot$.
The stellar mass range is determined to straddles the ``transition mass'' of $3\times10^{10}$ M$_\odot$ between blue cloud and red sequence galaxies.
This is because that one of the main goal of GASS project is to identify and quantify the incidence of galaxies that are transitioning between the blue, star-forming cloud and the red sequence of passively evolving galaxies (``transition galaxies'').
Here, main seven scientific outputs of the project are introduced:
\begin{itemize}

\item H\emissiontype{I} gas mass fraction \citep{2010MNRAS.403..683C,2012A&A...544A..65C}:
Cantinella et al. suggested that candidates of ``transition galaxies'' can be identified as outliers from the mean scaling relations between H\emissiontype{I} gas mass fraction ($M_{\rm HI}/M_\star$, where $M_{\rm HI}$ is H\emissiontype{I} gas mass) and other galaxy properties.
They found a scaling relation of galaxies among H\emissiontype{I} gas mass $M_{\rm HI}$, surface density of stellar mass $\mu_\star$, and $NUV-r$ color and interesting outliers from this plane:
gas-rich red sequence galaxies that may be in the process of regrowing their discs, as well as blue, but gas-poor spirals.


\item Star formation efficiency \citep{2010MNRAS.408..919S}:
Schiminovich et al. suggested candidates of ``transition galaxies'' are identified as (1) H\emissiontype{I}-rich but low-SFR and (2) H\emissiontype{I}-poor but high-SFR galaxies.
They explored the global scaling relations associated with the bin-averaged ratio of the SFR over the H\emissiontype{I} mass (i.e. $\Sigma_{\rm SFR}/\Sigma_{M_{\rm HI}}$ ), which is the H\emissiontype{I}-based star formation efficiency (SFE). 
They found that the H\emissiontype{I}-based SFE remains relatively constant across the sample with a value of $10^{-9.5}$ yr$^{-1}$ (or an equivalent gas consumption time-scale of $\sim3\times10^9$ yr) while 10~\% of galaxies have higher or lower SFE than the average value.
Such outliers are expected to have potential for a change (either decrease or increase) in their sSFR in the near future.

\item Inside-out formation of Galactic discs \citep{2011MNRAS.412.1081W}:
Wang et al. found an observational evidence indicating ``inside-out'' formation of galactic disks.
They showed that galaxies with larger H\emissiontype{I} fraction have bluer, more actively star-forming outer disks compared to the inner part of the galaxy.
They also found H\emissiontype{I} fraction and galaxy asymmetry do not have intrinsic connection, which suggests that gas is accreted smoothly on to the outer disc.

\item Baryonic mass-velocity-size relations \citep{2012MNRAS.420.1959C}:
Cantinella et al. found a generalized baryonic Faber-Jackson (BFJ) relation that holds for all the galaxies in their sample, regardless of morphology, inclination or gas content, and has a scatter smaller than 0.1~dex.
They compared baryonic Tully-Fisher (BTF) and BFJ relations for GASS sample, and investigated how galaxies scatter around the best fits obtained for subsets of disk-dominated and bulge-dominated systems.
They demonstrated that by applying a simple correction to the stellar velocity dispersions that depends only on the concentration index of the galaxy, disks and spheroids distribute on to the same dynamical relation.
This generalized BFJ relation indicates that there is a fundamental correlation between the global dark matter and baryonic content of galaxies regardless of morphology.

\item H\emissiontype{I} content and metal enrichment at galactic outskirts \citep{2012ApJ...745...66M}:
Moran et al. found that H\emissiontype{I}-rich galaxies tend to have larger metallicity drops in their outer disks, suggesting the accretion or radial transport of relatively pristine gas from beyond the galaxies' stellar disks.
They investigated radial metallicity profile of 174 star-forming galaxies using long-slit spectra and found that $\sim10$~\% of their sample exhibit a sharp downturn in metallicity.
They also found that the magnitude of the outer metallicity drop is well correlated with the total H\emissiontype{I} content of the galaxy.

\item Bivariate neutral hydrogen-stellar mass function \citep{2013ApJ...776...74L}:
Lemonias et al. investigated the bivariate neutral atomic hydrogen (H\emissiontype{I})-stellar mass function (HISMF) $\phi(M_{\rm HI}, M_\star)$ for 480 GASS galaxies. 
Based on the fitting to HISMF with Schechter function, they found that the slope of the HISMF at moderate H\emissiontype{I} masses correlate with stellar mass and SFR, whereas the characteristic H\emissiontype{I} mass varies weakly with them.

\item Environmental effect \citep{2013MNRAS.436...34C}:
Cantinella et al. claimed the importance of stripping of the cold interstellar medium in galaxy groups for galaxy evolution, which has not been considered in semi-analytic models of galaxy formation.
They found that massive galaxies located in haloes with masses of $10^{13}-10^{14}$ M$_\odot$ have at least 0.4~dex less H\emissiontype{I} than objects in lower density environments.

\end{itemize}

\subsubsection{Blind Ultra Deep H\emissiontype{I} Survey \citep[BUDHIES,]{2007ApJ...668L...9V}}

BUDHIES is a deep H\emissiontype{I} survey of galaxies in two clusters in intermediate redshifts, Abell~2192 ($z=0.187$) and Abell~963 ($z=0.206$), and the large-scale structure of the Universe around them with WSRT.
Abell~963 is a massive lensing Butcher-Oemler \citep{1978ApJ...219...18B} cluster with a large fraction of blue galaxies \citep[$f_{\rm B}=0.19$,][]{1983ApJS...52..183B}, and a total X-ray luminosity of $L_X \simeq 3.4 \pm 1 \times 10^{44} h^{-2}$ erg s$^{-1}$ \citep{2003MNRAS.342..287A}.
Abell~2192 is a less massive cluster in the process of forming, with a high degree of substructure \citep{2012ApJ...756L..28J}.
This cluster is barely detected in X-rays \citep[$L_X \simeq 7\times 10^{43} h^{-2}$ erg s$^{-1}$;][]{1999A&A...349..389V}.

The survey aims at understanding where, how, and why star-forming spiral galaxies get transformed into passive early-type galaxies.
In this project, H\emissiontype{I} emissions were successfully detected in 127 galaxies of Abell~963 and 36 of Abell~2192 with H\emissiontype{I} masses of $2\times10^9$ M$_\odot$.
In Abell~2192, Jaff{\'e} et al. found that the incidence of H\emissiontype{I}-detections significantly correlates with environment \citep{2012ApJ...756L..28J,2013MNRAS.431.2111J}:
at large clustercentric radii ($>2-3$ virial radii), many galaxies are detected in H\emissiontype{I}, while at the core of the forming cluster, none of the galaxies are H\emissiontype{I}-detected.
They found that this effect starts to become significant in low-mass groups that pre-process the galaxies before they enter the cluster, so that by the time the group galaxies fall into the cluster they may already be H\emissiontype{I} deficient or even devoid of H\emissiontype{I}.

\subsubsection{COSMOS H\emissiontype{I} Large Extragalactic Survey \citep[CHILES,][]{2013ApJ...770L..29F}}

CHILES is a 1000-hrs Very Large Array (VLA) project to produce the first H\emissiontype{I} deep field.
The observations are carried out with the VLA in B configuration and cover a redshift range of $0<z<0.45$.
The field is centered at the COSMOS field \citep[Cosmic Evolution Survey,][]{2007ApJS..172....1S}, which has been already observed in from radio to X-ray.
This project will provide H\emissiontype{I} images of at least 300 galaxies spread over the entire redshift range to investigate the following topics:
1) H\emissiontype{I} content, morphology and kinematics of individual galaxies, 2) H\emissiontype{I} mass function, 3) Cosmic Neutral Gas Density, 4) Environmental suppression of star formation, and 5) Dark haloes and baryons.
\citet{2013ApJ...770L..29F} presented initial results based on 50-hrs observations.
They detected H\emissiontype{I} emissions from 33 galaxies within $34'\times34'$ field for now, including three without a previously known spectroscopic redshift.
The highest redshift of galaxies with H\emissiontype{I} measurement is $z=0.176$.


\subsection{Toward higher-redshifts: revealing galaxy evolution through H\emissiontype{I}
absorption line systems}\label{subsec:world_HI}

In observing distant galaxies, because of limited
sensitivities of observational facilities,
the sample is always biased to the brightest populations.
Moreover, since the emission from a galaxy is often related to
its star formation activity, it is relatively easy to detect
a growing galaxy population whose star formation rate is high or
a grown galaxy population which has already built up its
stellar mass. This means that the observations of the
emission from distant or high-redshift (high-$z$) galaxies tend to miss
ones which have not yet converted a significant fraction of
the ISM to stars.

Such galaxies with gas-rich and low-stellar-mass properties
can be sampled if we directly sample galaxies with its gas
content. The most common method of sampling this kind of
galaxies is to observe gas absorption lines in bright continuum
spectra of background quasars [quasi-stellar objects (QSOs)].
In particular, Ly$\alpha$ absorption in the continuum spectra
of QSOs is useful to identify the foreground (intervening)
neutral hydrogen (H\emissiontype{I}) in the ISM and the intergalactic
medium (IGM) over a
wide column density range. A sample taken by this method
is called Ly$\alpha$ absorption-line systems or Ly$\alpha$ forest
\citep{sargent80,smith86,wolfe86}.

Damped Lyman $\alpha$ systems (DLAs) are the QSO
absorption-line systems whose H\,{\footnotesize I} column 
density is higher than $2 \times 10^{20}~\mathrm{cm^{-2}}$
\citep{prochaska00}.  Because of the
prominent  Ly$\alpha$ absorption in bright QSO continuum,
DLAs provide us with unique opportunities to 
trace high-$z$ galaxy evolution. 
The statistics of H\,{\footnotesize I} column density 
in DLAs is expected to reflect the total H\,{\footnotesize I} 
abundance contained in high-$z$ objects \citep{peroux03}
and the density structure 
of the interstellar medium (ISM) in them \citep{razoumov06}. 
 
\begin{figure}
\begin{center}
\includegraphics[width=0.45\textwidth]{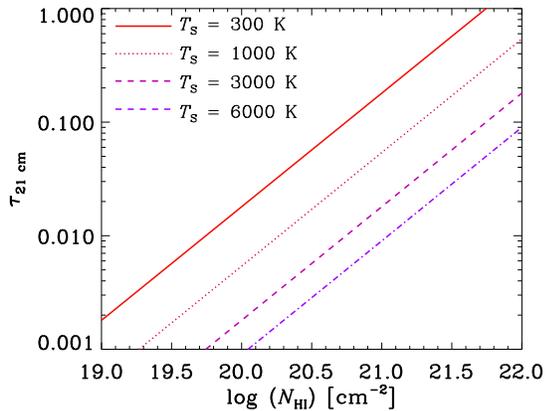}
\end{center}
\caption{H\emissiontype{I} 21 cm absorption optical depth
$\tau_\mathrm{21\,cm}$
as a function of H\emissiontype{I} column density $N_\mathrm{HI}$
with various spin temperatures ($T_\mathrm{s}=300$,
1,000, 3,000, and 6,000 K for the solid, dotted, dashed,
and dot-dashed lines, respectively). The velocity dispersion
is fixed as $\Delta v=10$ km s$^{-1}$.}
\label{fig:tau21_NHI}
\end{figure}

The hyperfine transition of H\,{\footnotesize I} at a wavelength of
21~cm is also known to be a good tracer of H\,{\footnotesize I}.
While the Ly$\alpha$ absorption is determined by
the H\,{\footnotesize I} column  density, $N_{\mathrm{HI}}$,
along the line of sight, 
the 21~cm optical depth, $\tau_{\mathrm{21\,cm}}$, depends on 
the spin temperature as well \citep[e.g.,][]{furlanetto06}:
\begin{eqnarray}
\tau_\mathrm{21\,cm} = 0.054
\left(\frac{T_\mathrm{s}}{10^3\,\mathrm{K}}\right)^{-1}
\left(\frac{\Delta v}{10\,\mathrm{km\,s^{-1}}}\right)^{-1}
\left(\frac{N_\mathrm{HI}}{10^{21}\,\mathrm{cm^{-2}} }\right) , \nonumber \\
\label{eq:tau21}
\end{eqnarray}
where $T_\mathrm{s}$ is the spin temperature of hydrogen atoms
and $\Delta v$ is the velocity dispersion.
Therefore, combination of Ly$\alpha$ and 21 cm 
optical depths enables us to derive both column 
density and temperature. In Figure \ref{fig:tau21_NHI},
we show $\tau_\mathrm{21\,cm}$ as a function of $N_\mathrm{HI}$
for various $T_\mathrm{s}$ with $\Delta v=10$ km s$^{-1}$.

There have been some attempts to detect 21 cm line absorption in
the radio continuum of DLA-bearing quasars. The fraction of DLAs
with detection of 21 cm absorption is, however, low
\citep[e.g.][]{curran10}. A systematic 21 cm absorption surveys of
DLAs has been carried out by \citet{srianand12} using
the Giant Metrewave Radio Telescope (GMRT) and
the Green Bank Telescope (GBT), but 21 cm absorption is
detected for only one out of ten DLAs at
$2<z_\mathrm{{abs}}< 3.4$ ($z_\mathrm{{abs}}$ is 
the absorption redshift of the DLA). They further included samples in the literature, and
interpreted the small fraction of 21 cm absorption detection as a
large covering fraction of the warm neutral medium with
temperature $\gtsim 10^3$ K in the DLA host galaxies
\citep[see also][]{kanekar13,kanekar14}.
Even for the detected cases, the derived spin temperatures are
mostly high ($\sim 10^3$~K)
\citep[see also][]{wolfe79,wolfe85,debruyn96,carilli96,briggs97,kanekar06}.

It has been pointed out that DLAs may be biased to
systems with low dust extinction: if a foreground
system has high dust extinction, the ultraviolet (UV)
continuum of
the background QSO can become too faint to be detected
by optical telescopes \citep{fall93}. The detection rate of
high-$N_\mathrm{HI}$
DLAs may be significantly suppressed by dust extinction
\citep{vladilo05}.
This bias could be avoided if we sample background QSOs
at radio frequencies, where the effect of dust extinction
is negligible. In fact, systematic studies of DLAs for
radio-selected QSOs have been performed
\citep[e.g.,][]{akerman05,ellison05}.
Although these samples did not show the presence of
a significant bias caused by dust extinction,
the sample size is not large enough.
In particular, the number of systems rapidly declines
as $N_\mathrm{HI}$ becomes large. Sampling of a large
number of QSO in the radio will be one of the most
important goal of SKA surveys in the future.
With a large-area ($\sim$10,000 deg$^2$) survey with SKA1-MID proposed
by \citet{morganti15}, 5$\sigma$ detection of typical
DLAs with $\tau_\mathrm{21\,cm}\sim 0.015$ would be
possible against background sources $>30$ mJy
($\tau_\mathrm{21\,cm}\sim 0.05$ against $\sim$10 mJy and
$\tau_\mathrm{21\,cm}\sim 0.1$ against $\sim$2--3 mJy)
up to $z\sim 3$, which is limited by the lowest frequency
of SKA-MID (350 MHz). Surveys with SKA1-LOW would extend the
redshift range towards higher $z$.

Before SKA, the following pathfinder observations are
planned \citep{morganti15}. The Australian Square Kilometre
Array Pathfinder (ASKAP) has a wide field of view
(FoV $\sim 30$ deg$^2$) and frequency coverage down to
700 MHz. The ASKAP First Large Absorption Survey in H\emissiontype{I}
(FLASH) will be able to detect bright QSOs and radio galaxies,
and will open up systematic studies of intervening
21 cm absorption at $0.5<z<1$. With the planned
integration time of 2 hours per field, the 5 $\sigma$ detection
limit for the 21 cm absorption optical depth is
$\tau_\mathrm{21\,cm}\sim 0.01$ for 1 Jy
background sources while it is $\tau_\mathrm{21\,cm}\sim 0.3$
for 50 mJy sources. Apertif and MeerKAT will achieve deeper
$\tau_\mathrm{21\,cm}$ limits with smaller survey areas.

\subsection{Galaxy evolution history through radio continuum}
\label{subsec:world_conti}

Radio continuum emission from galaxies is composed of
thermal free--free radiation from H\emissiontype{II}
regions and nonthermal synchrotron radiation from
supernova remnants \citep{condon92} (for the simplicity of
discussion, we
do not consider radio emission from AGNs in this section).
Since both of these emission sources
are related to massive stars, the luminosity of
radio continuum is known to be a good indicator of
star formation rate \citep{condon92}.
Unlike UV--optical star formation indicators,
radio continuum is not affected by dust extinction,
so that no correction for dust extinction is required in the
conversion from radio continuum luminosity to star
formation rate.
If we detect radio continuum emission from galaxies
up to high redshifts, we can in principle clarify the cosmic
star formation history.

Another extinction-free star formation indicator is
far-infrared (FIR) continuum luminosity
\citep{kennicutt98,inoue00}. It is known that radio
and FIR continuum luminosities are tightly correlated
\citep{condon92,yun01}. This radio--FIR correlation holds
very well for nearby star-forming galaxies. However,
whether or not galaxies keep the radio--FIR correlation
over their evolution is not obvious. It is rather expected
that the radio--FIR relation changes as galaxies evolve,
since FIR emission depends on the dust content and
radio emission is affected by interstellar magnetic field
strength and production/diffusion of high-energy electrons.
The different time-scales of
these processes may indicate a possibility that
the radio--FIR relation changes at high redshifts.
It has been clarified observationally that the radio--FIR
hold with a factor of 2--3 times smaller FIR/radio ratio
at $z\ltsim 2$
\citep[e.g.,][]{garrett02,gruppioni03,ibar08,murphy09}.
\citet{michalowski10} reported a similar relation to
$z\sim 2$ holds even at $z\sim 5$.

The understanding of dust emission in high-redshift galaxies
is rapidly progressing because of advanced observational
facilities such as \textit{Herschel}
\citep[e.g.,][]{rowlands14}. ALMA, with its unprecedented
sensitivity, is starting to reveal the high-redshift
star formation activities through dust emission.
However, the narrow FoV of ALMA makes the survey efficiency
low. At
radio wavelengths, on the other hand,
although some surveys are planned
using the Jansky Very Large Array (JVLA),
future SKA observations are crucial to detect
``normal'' star-forming galaxies whose star
formation rate is $\sim 10$ M$_\odot$ yr$^{-1}$
\citep{murphy09,jarvis15}.
The wide FoV of SKA will provide a
unique opportunity of tracing the cosmic star formation
history at wavelengths free from dust extinction.

\citet{murphy15} focused on the ultra-deep SKA1-MID/Band 5
reference survey \citep{prandoni15} at a frequency range of
4.6--13.8 GHz. Most previous surveys targeted
low frequencies such as 1.4 GHz because FoVs are wider
and galaxies are brighter. For example,
\citet{seymour08} derived the cosmic star formation history
at $z<3$ using Multi-Element Radio-LInked Network (MERLIN)
and the Very Large Array (VLA) at 1.4 GHz as well as
the VLA at 4.8 GHz, obtaining a consistent star formation
history with various optical surveys
\citep[see also][]{morrison10}.
\citet{smolcic09} derived the cosmic star formation
history and the evolution of radio luminosity function
out to $z=1.3$ by observing the COSMOS field with the VLA
at 1.4 GHz.
\citet{smolcic14} observed the same field at an even
lower frequencies (324 MHz) by the VLA, and found
objects with steep spectral slopes at $z\gtsim 1$.

Compared to those surveys, \citet{murphy15} focus on
higher frequencies for SKA, since the high sensitivity of
SKA enables us to perform a survey deep enough even at
frequencies $\gtsim 10$ GHz. According to their expectation,
about 30 and 85 sources per arcmin$^2$ are detected.
Such a high frequency has an advantage of high angular
resolution. If we can take a 200 km baseline,
an angular resolution of $<0.03$ arcsec can be achieved
at $>$10 GHz (yet with enough short-baseline data
to prevent galaxies from being resolved out),
which means that we can resolve 250~pc
at $z\gtsim 1$. This angular resolution matches what
is achieved by future space optical--near-infrared
telescopes such as \textit{James Webb Space Telescope}
(\textit{JWST})\footnote{http://www.jwst.nasa.gov/}.
Another advantage of high-frequency surveys is that
the contribution from free--free emission is relatively
large compared to low frequencies (such as 1.4 GHz).
At high redshift, it is expected that high-energy
electrons lose their energy through inverse Compton
scattering of the cosmic background radiation, which
means that synchrotron emission may not be a good
indicator of the star formation activities at high
redshift \citep{murphy09}.

\begin{figure}
\begin{center}
\includegraphics[width=8cm]{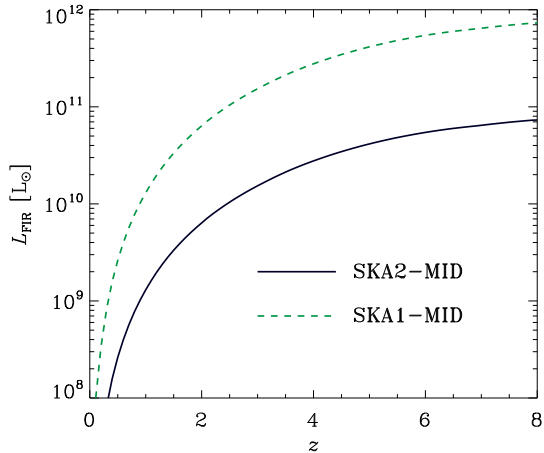}
\end{center}
\caption{The total FIR luminosity corresponding to the
5$\sigma$ detection limits of SKA2-MID and SKA1-MID (solid
and dashed lines, respectively), as a function of redshift.}
\label{fig:detection_LFIR} 
\end{figure}

It is worth mentioning the comparison with
Next Generation Very Large Array (ngVLA)
\citep{casey15}.
The ngVLA will achieve a factor of 5--10 times
improvement over the current JVLA. By assuming
10 times better sensitivity than VLA, ngVLA
it will achieve a 5$\sigma$ detection limit of
0.06 $\mu$Jy for 1000 hr of
integration time (we multiplied 0.1 to the detection
limit estimated for JVLA by \citealt{murphy15}) at 10 GHz.
As cited in Section \ref{subsubsec:survey},
SKA1-MID and SKA2-MID will achieve
a 5 $\sigma$ detection limit of 0.2 and 0.02 $\mu$Jy,
respectively, with 1000 hr of integration .
Therefore, ngVLA will have
an intermediate capability between SKA1-MID and SKA2-MID.
For convenience, we show the detection limit at various
redshifts in Fig.\ \ref{fig:detection_LFIR}, where the galaxy
luminosity is indicated by the far-infrared (FIR) luminosity
(the total luminosity emitted by dust, which could be compared to the
8--1000 $\mu$m luminosity). The FIR--radio correlation
in \citet{totani02} is used for the relation between FIR and
radio luminosities (see Section \ref{subsubsec:survey} for details).

Although we emphasized high-frequency surveys above,
we must mention that tracing both free--free
and synchrotron components is the strong point of
the wide frequency coverage of SKA.
The synchrotron component provides independent
information from the free--free component, since
it depends on the magnetic field strength in the ISM.
Therefore, decomposing those two components serves as
a tracer of not only star formation but also
amplification of magnetic field.

\section{Studies by SKA-Japan Members}\label{sec:SKA_JP}

We introduce the strong point in the studies of galaxies in Japan. 
Observationally the rich and deep knowledge and experience of radio astronomy 
accumulated by Nobeyama Radio Observatory etc. 
Especially we have a practical advantage to have an access to the ALMA observation time. 
As being very well-known, ALMA is suitable for observing emissions from dust and 
various molecules.
The combination of such observations and HI data from SKA, we can obtain a 
global picture of the recycling of the ISM from atomic gas to the SF and
metal enrichment via molecular gas. 
As for the theoretical side, we have a strong point for the simulation of galaxy formation
and chemical evolution theory.
Thanks to this, as well as theoretical predictions, we can also make a quick comparison when 
less luminous and/or more distant galaxies are detected by SKA and ALMA. 

For the studies of galaxy formation and evolution, we consider three representative
redshift ranges.
First is the early Universe, in which stars are formed from gas and galaxies grow. namely
the nepionic period of galaxies, at $z > 3$. 
In particular, direct observation of a system with an enormous amount of gas just
before the first starburst is crucial to examine the physics of galaxy formation. 
Second is the epoch in which galaxies establish their global properties through 
the hierarchical structure formation at $1 < z < 2$. 
Galaxy morphology emerges in this period.
Also the average SFR, fraction of SF activity hidden by dust, and merging rate of 
dark halos and galaxies peak around this redshift range.
Then this is a tumultuous time of galaxy evolution. 
The last but not least, the important epoch is $z = 0$.
Though it looks paradoxical, the {\it evolution} can be discussed only when the zero-point,
the property of nearby galaxies, is well defined. 
The Local Universe is also important to make a crossover of the ISM physics and 
extragalactic physics, to construct a large flow of ISM--nearby galaxies--high-$z$
galaxies.
We introduce our scientific themes related to these three epochs in the following.

\subsection{Nearby Galaxies: Evolution and Characteristics of the ISM and Unified Understanding of Star Formation}

\begin{figure*}[th]
  \begin{center}
   \includegraphics[width=12cm]{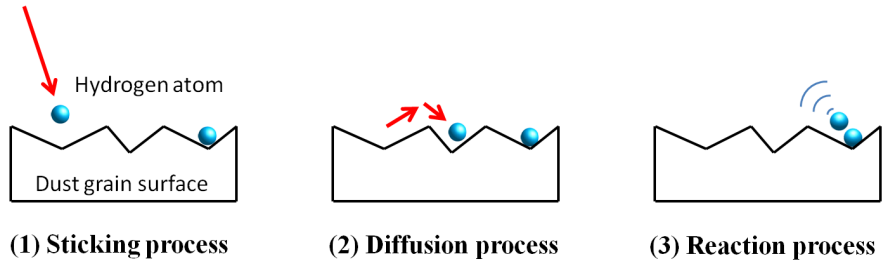} 
  \end{center}
 \caption{The four processes for molecular gas formation on dust particles 
\citep[adopted from][]{takahashi2000}. }
\label{fig1}
\end{figure*}

Here we describe the science targets that Japan will pursue using SKA in the subject of 
nearby ($\sim 10$~Mpc) galaxy studies.  In particular, we focus on 
(1) the understanding of "dark gas", which is believed to be the dominant component in the transition phase between the atomic and 
molecular gas, and (2) formation of stars from atomic gas in low metallicity environments.  The discussion 
provided here is naturally linked to the SKA-JP ISM subgroup.

\subsubsection{Dark Gas} 

The 21 cm line has been used as a gas mass tracer under the assumption that the line is optically thin.  
Contrary to this assumption, \citet{fukui15} suggests the presence of optically thick atomic gas in 
about 50\% of the gas traced in 21 cm observations, based on their detailed comparison between the 
distribution of dust (from Planck) and H\emissiontype{I} data.  They further suggest that this optically thick gas is the 
component that dominates the density range of 100 - 1000 cm$^{-3}$, and undetected in 
CO or 21 cm observations ("dark gas").     The important future direction is to compare the 
amount/distribution/kinematics of the ionized/atomic/molecular gas (including dark gas) 
in galactic and extra galactic sources.  Pioneering studies have been conducted using 
the Nobeyama 45m telescope \citep[e.g.][]{nakanishi06, tosaki11, kaneko13},
The Milky Way galaxy is the best target for a detailed comparison 
between the galactic structure and the characteristics of the ISM, but our location in the Galaxy
prevents us to conduct such a detailed study.  The large scale structure such as spiral arms or 
bar structure can have a significant effect on the physical condition of the galactic ISM.  SKA will 
allow us to study a typical grand-design spiral galaxy located at 10~Mpc using $\sim 0.1$ arcsec (5~pc)
angular resolution.  An important science synergy is to compare the atomic gas distribution 
obtained with SKA, and the dust distribution obtained with ALMA in nearby spiral galaxies.

The metallicity, which is proportional to the amount of dust, affects the physical condition of 
the ISM in galaxies, in addition to the kinematical disturbance mentioned above.  From recent theoretical studies, 
it is found that the molecular ISM is not the necessary condition for star formation, but the atomic 
gas can also lead to star formation in low metallically environments.  
Therefore, detailed understanding of the atomic gas in the diffuse outskirts of galaxies or dwarf galaxies 
can be one of the science goals of SKA.  We summarize the current understanding of the 
atomic to molecular transition and the associated star formation, following the review provided by
\citet{krumholz13},  in what follows.

\subsubsection{Formation of Stars From Atomic Gas in Low Metallicity Environments} 

In nearby galaxies, the star formation rate is well correlated with molecular gas, 
but the correlation with atomic gas is rather poor \citep{wb02, kennicutt07, bigiel08, leroy08}.
 The gas depletion time scale $t_{\rm dep}$,
which is the ratio between molecular gas mass and star formation rate, is nearly constant 
in nearby galaxies ($\sim 2$~Gyr; \cite{bigiel11}).

The balance between molecular gas formation and dissociation determines the 
chemical condition of the interstellar hydrogen \citep{omukai12}.  Molecular gas formation 
occurs on the surface of dust if the metallically is higher than 1/100,000 of the solar value.
The dissociation rate under a typical solar UV radiation is $5 \times 10^{-11}$ s$^{-1}$ 
per one hydrogen molecule \citep{draine96}, while the formation rate under 
a molecular hydrogen column of 100~cm$^{-3}$ is $3 \times 10^{-15}$ s$^{-1}$ 
per one hydrogen molecule.  Therefore, the predicted molecular hydrogen abundance is 
only $10^{-4}$, if this formation and dissociation balance is maintained.  In reality, 
the Lyman-Werner UV photons are efficiently shielded by the presence of dust, allowing 
molecular clouds to form.  Thus atomic hydrogen dominates the exterior region where the effect of dust 
attenuation is negligible, while the inner regions are more dominated by the molecular gas due to the
abundance of dust \cite[e.g.,][]{vdb86,sternberg88,ns96,liszt02,gml07,krumholz08,gnedin09, mk10, mlg12}.

The chemical and thermal condition of the interstellar hydrogen gas are both related to the 
shielding of UV light due to interstellar dust, and this may be the reason for the good correlation 
between molecular gas and star formation.  The chemical condition is determined by the 
balance between molecular hydrogen formation on dust particles and the photo dissociation 
due to UV photons.  The thermal condition is determined by the balance between the 
combination of photoelectric heating due to UV photons and the thermal exchange of dust/gas 
and the cooling due to line emission after collisional excitation.  Therefore, the 
dependence of thermal/chemical process to the volume/column density and the UV radiation field 
are extremely similar.  The cooling rate and the formation rate are both proportional 
to the density and metallically ($n^2 Z'$), because the cooling due to line emission and 
the formation of molecular hydrogen are both due to collisions.  The UV dissociation and photoelectric 
heating are both dependent on the UV radiation field and dust attenuation in the same way.

Because of this similarity, the correlation holds well in a wide range of gas densities, dust attenuation, 
and metallicity. The transition from atomic to molecular gas occurs at $>$ 100~K to 10~K, and this 
is determined by the attenuation of dust: the temperature is low in regions where dust attenuation
is efficient whereas the temperature is high in regions with little dust attenuation.  Therefore, 
molecular hydrogen and star formation show a good correlation: the transition from atomic to molecular 
hydrogen is not directly related to the triggering of star formation, but it is rather related to 
the decrease in gas temperature.

In low metallicity environments, the timescale for the transition from atomic to molecular hydrogen is comparable
to the dynamical timescale, but the timescale to reach thermodynamical equilibrium is 1/1000 shorter.  
The chemical and thermal timescales both depend on the metallicity, and they are longer in low metallicity
environments.  On the other hand, the timescale for star formation does not depend on metallicity.   
Therefore there must exist a certain value of metallicity that satisfies a condition in which (thermal timescale) $<$ 
(dynamical timescale) $<$ 
(chemical timescale).  This suggests that star formation occurs before interstellar hydrogen becomes molecular,
if the process that governs star formation is indeed the chemical condition and not the thermal condition.
In such a condition, the star formation rate should correlate better with atomic hydrogen rather than atomic 
hydrogen.  This is also suggested in the numerical simulations by \citet{gk12}, and further 
suggested in the analytical model by \citet{krumholz12} that this effect is seen in the ISM with metallicities of
1-10\% of the solar value.  

The SKA-JP galaxy evolution subgroup will collaborate with the ALMA group, and investigate the 
spatial distribution of the atomic-molecular transition as a function of metallicity and galaxy evolution stage 
with a future goal to compare the physical processes with the galaxies in the high-z universe.


 

\subsection{Construction and understanding of the extended scaling laws of galaxies connecting
gas, dust, and SF}

Once the main body of a galaxy has established, a tight relation between some
global properties of galaxies emerges.
This is referred to as the scaling relation, playing an important role to study 
the evolution of galaxies. 
The best-known scaling relation may be the above mentioned TF and BTF relations.
There are also many other scaling relations. 
However, we should note that most of them are heuristic relations without physical
explanation. 
Here we introduce some scaling relations tightly connected to the galaxy evolution, 
and discuss the way to a synthesis and theoretial reasoning of these relations.

As we mentioned, the SFR is the most interested physical quantity from the point of
view of galaxy evolution. 
In 2015, the relation that gathers researchers' attention is the so-called ``the main 
sequence of SF galaxies'' .
This is a tight linear sequence of SF galaxies on the (logarithms of) stellar mass $M_*$--SFR plane.
Historically this is an equivalent to the color--magnitude relation, but since it is presented 
with more physical quantities like $M_*$ and SFR, the scaling relation of SF galaxies 
appears very prominently \citep[e.g.,][among others]{schiminovich2007}. 
It is not surprising that a trivial scaling relation exists since both are global properties
of galaxies. 
What we should pay attention is that the slope of the linear relation is shallower than
unity, meaning the more massive a galaxy is, the less efficient the SF activity becomes.
This trend is referred to as the downsizing, which was recognized in late 1980's but 
has not been physically understood until now.


Starburst galaxies like ultraluminous infrared galaxies (ULIRGs) do not locate on
the main sequence, but distribute more than an order of magnitude above it
\citep[e.g., ][]{buat2005,buat2007}.
This implies that the main sequence is the relation of quiescently evolving 
SF galaxies, or the staying time is long in the sequence.
Galaxies that already stopped their SF has $\mbox{SFR} \sim 0$ and do not
appear on this plane. 
Therefore, studies from various point of view are necessary to understand 
the relation.
\citet{genzel2012} examined the amount of molecular gas in the SF main 
sequence galaxies.
The redshifts of their sample is at $0 < z < 2$, but it is not yet a systematic statistical
study. 
\citet{magnelli2012} and \citet{magnelli2014} examined the relation between 
the temperatures of dust and molecular gas in the main sequence galaxies.
However since the observational data of H{\sc i} is limited to $z < 0.5$ at this moment,
it is difficult to discuss its evolutionary aspect. 
Also in relation between the main sequence and the Kennicutt--Schmidt law
discussed in the following, a spatially resolved analysis of galaxies is necessary
\citep[e.g., ][]{teruya2014}. 
Studies to reveal the mechanism which shapes the main sequence of SF galaxies 
in terms of the ISM physics will only be possible with ALMA and SKA.

The SF is mainly determined by relatively local conditions of the ISM
in galaxies
However, it has been well known that the SF and integrated properties of 
a galaxy are correlated \citep[e.g., ][]{roberts1994, kennicutt2012}.
The existence of a correlation between local properties, like the spiral arms, molecular clouds, 
H{\sc ii} regions, etc., and global properties like the morphology, total mass, etc., is not
a trivial phenomenon at all but an important problem to be solved by the astrophysics. 
A relation between the (surface) densities of gas and SFR, which appears in various
spatial scales, is the Kennicutt--Schmidt (KS) law \citep[see e.g., ][as a review]{kennicutt2012}.
The KS law is expressed as follows: 
\begin{eqnarray}
  \Sigma_{\rm SFR} \propto \Sigma_{\rm gas}^n \; .
\end{eqnarray}
This relation is a single power law for more than six orders of magnitude.
The index $n$ is observationally in a  range of $1 < n < 2$. 
Whether $n$ is closer to 1 or 2 determines the plausible scenario of the physical
process in a molecular gas to form stars \citep[e.g., ][]{momose2013}.
The KS law is an important ingredient in the theory of chemical evolution in
galaxies, and gives a great impact if it will be physically understood. 
Thus, the KS law is of great interests from ISM to the extragalactic physics. 

However, we should point out that still the KS law is a purely empirical relation.
Further, the debate on the index $n$ is not yet converged. 
In order to derive the KS law from physical processes like the gravitational collapse
and/or collision of molecular clouds, a large galaxy sample with neutral and molecular
gas information selected by a homogeneous criterion 
is necessary to present the relation with errors as small as possible. 
SKA will play a unique role utterly impossible with other instruments.
With the sample obtained by SKA, we can explore the spatially resolved KS law
up to $z \sim 1$ and clarify the governing physical processes.
By synthesizing the all scaling laws, we will find a unified scaling law of galaxies.
We try to derive this unified relation from a first principle. 
Studies on galaxy evolution will be mature by such an approach.

\subsection{Exploration of galaxy evolution through H\emissiontype{I} 21-cm absorption lines}

As discussed in Section \ref{subsec:world_HI},
gas absorption in the lines of sight of bright continuum
sources such as QSOs will enable us to trace the ISM and
IGM at various redshifts. In particular, Ly$\alpha$ absorption
line systems are often used for such studies.
There have been some attempts of detecting
21 cm absorption in DLAs
(Ly$\alpha$ absorption systems with
$N_\mathrm{HI}>2\times 10^{20}$ cm$^{-2}$), but
most of the DLAs are not detected, indicating
small $\tau_\mathrm{21\,cm}$. This suggests that
a large fraction of the area of the ISM in the
intervening systems is covered by the warm ISM,
whose typical temperature is higher than 1,000 K
(equation \ref{eq:tau21}).

A high covering fraction of the warm ISM in
DLA hosts is also suggested by
\citet[][hereafter H03]{hirashita03}, who interpreted the small
fraction of
H$_\mathrm{2}$ (Lyman-Werner band absorption) detection 
for DLAs as due to the predominant warm medium in which
the equilibrium H$_\mathrm{2}$ fraction
($f_\mathrm{H_2}$: the fraction of hydrogen nuclei in the form of H$_\mathrm{2}$)
is as low as $\ltsim 10^{-6}$.
Indeed, from some observations, although 
the H$_\mathrm{2}$ fraction is largely enhanced for some DLAs, 
stringent upper limits  ($f_\mathrm{H_2}\ltsim 10^{-7}$--$10^{-5}$) are
laid on a significant fraction  of DLAs
\citep*{black87,petitjean00}.
This can also be interpreted as due to a low formation rate
of H$_\mathrm{2}$ in dust-poor environments relative to the Milky Way
\citep*{levshakov00,ledoux03,petitjean06}
or a high H$_\mathrm{2}$ dissociation 
rate by strong UV radiation \citep{petitjean00}.
However, we should keep in
mind that such upper limits do not exclude the existence of
molecule-rich clouds in these systems, because such
clouds may have a very low volume filling factor and may 
be rarely located in the line of sight as shown by our
study above (H03). Therefore, it appears
that the ISM structures, in which dense regions are 
localized while most of the volume is occupied by the 
warm diffuse medium, provide a common interpretation
for the lack of detections in both H\,{\footnotesize I} 21 cm
absorption and H$_{\mathrm{2}}$ in DLAs.

One of the possible contributions from the
SKA-Japan community is to provide a numerical simulation
scheme which can predict the statistics of 21 cm
line optical depth based on high-resolution density
and temperature structures and chemical reactions
(especially H$_2$ formation on dust).
High-resolution simulations of the ISM in the cosmological
structure formation scenario have been performed
by e.g., \citet{saitoh09}, \citet{okamoto13}, and \citet{yajima15}.
These simulations will enables us to obtain a tool for interpreting the
statistics of $\tau_\mathrm{21\,cm}$ through its dependence
on $N_\mathrm{HI}$ and $T_\mathrm{s}$ (equation \ref{eq:tau21}).
We could also implement chemical reactions of other
species such as CO based on the formulation by
\citet{inoue07}, who considered gamma-ray bursts instead of
QSOs for background sources.
Moreover, the treatment of H$_2$ formation on dust
could be refined based on recent efforts on implementing
dust enrichment and H$_2$ formation in galaxy evolution
simulations \citep{bekki15}.
Detailed modeling of dust enrichment by
\citet{asano13} could be implemented in the simulations.
These lines of modeling will enable us to make a unique
contribution to H {\footnotesize I} absorption studies by SKA.

In the following, we demonstrate that modeling of
the ISM based on a high-resolution simulation is useful
to interpret the 21 cm absorption statistics.
We use our previous 2-dimensional model in H03
for a first step, since it was used in our previous studies
to interpret the paucity of H$_2$ detection in DLAs.
The results will be used as a basis for a future development
using the state-of-the-art simulations mentioned above.
\citet{braun12}
adopted the observational H\,{\footnotesize I} maps of nearby
galaxies with a typical spatial resolution of 15--100 pc
and predicted the statistics of $\tau_\mathrm{21~cm}$ for
DLAs. However, we are also interested in the H$_2$ abundance,
and it is difficult to get observational H$_2$ map down to the level of
$f_\mathrm{H_2}\sim 10^{-6}$ (typical detection limit for DLAs)
even for nearby galaxies.
The theoretically made H$_2$ map by H03 can serve
to overcome this difficulty.
Since H$_2$ formation is related to the dust abundance
\citep[see e.g.][]{fynbo11},
we can also include
the bias in the optical selection
of QSOs caused by dust extinction in a self-consistent way.
We use the same values as in H03 for the cosmological parameters:
$(h,\,\Omega_\mathrm{m},\,\Omega_\Lambda ,\,\Omega_\mathrm{b})
=(0.7,\,0.3,\,0.7,\, 0.02h^{-2})$.


\subsubsection{Simulation of the ISM structures in a DLA host galaxy}
\label{subsubsec:simulation}

We briefly review the hydrodynamical simulation in H03.
It is a 2-dimensional hydrodynamical simulation of a galactic
disc based on \citet{wada01}.
We chose the parameter values
appropriate for formation 
redshift $z_\mathrm{{vir}}=3$ and
virial mass $M_\mathrm{{vir}}=8.0 \times 10^{10}$ M$_{\odot}$. 
A 1 kpc $\times$ 1 kpc area is simulated with 2048 $\times$ 2048 grids.
We refer the interested reader to H03 for
the simulated hydrogen column density 
($N_\mathrm{H}$) and gas temperature ($T_\mathrm{{\rm gas}}$) maps.
Although they concentrate 
on H$_\mathrm{2}$ fraction in DLAs, the same framework is applicable
for 21 cm optical depth.
To avoid the effect 
of the boundary conditions we only use the central
450 pc radius. Although the host galaxies of DLAs are
not necessarily large disk galaxies, we only use the 
simulation to model the structures of 
the ISM created by hydrodynamical evolution. We assume 
that such hydrodynamical structures are common for possible DLA hosts.




To choose the possible lines of sight for DLAs, we only adopt the grids 
where $N_\mathrm{H}\geq 2\times 10^{20} \mathrm{cm}^{-2}$. We also
introduce the dust extinction effect, since if the extinction is so large
that the background QSO is obscured, such a line of sight cannot sample DLAs. 
\citet{vladilo05} showed strong lack of 
DLAs with zinc column density $N_\mathrm{Zn}>10^{13.2} \mathrm{cm}^{-2}$,
which they explained with 
the effect of extinction in metal-rich environments.
For each of the grids in our simulated galaxy, 
we calculate $N_\mathrm{Zn}$ as
$N_\mathrm{Zn}=10^{-7.35} ({Z}/{\mathrm{Z}_{\odot}})N_\mathrm{H}$,
where the factor $10^{-7.35}$ is the number ratio of Zn to H in the solar
composition \citep{anders89} and ${Z}/\mathrm{Z}_{\odot}$ is
the metallicity normalized
to the solar metallicity. We treat the metallicity as a free parameter.
We exclude those grids where  $N_\mathrm{Zn}>10^{13.2}~\mathrm{cm^{-2}}$,
consistent with the observed extinction bias mentioned above.



We calculate the optical depth of 21 cm absorption
$\tau_\mathrm{{21~cm}}$ on each grid using equation (\ref{eq:tau21}).
 We assume that 
$T_\mathrm{s}=T_\mathrm{{\rm gas}}$, since the density is high enough for 
the spin temperature to approach the kinetic temperature
\citep{field58}.
We also assume that all hydrogen is in atomic form;
i.e.\ $N_\mathrm{HI}=N_\mathrm{H}$.
We adopt $\Delta v=10~\mathrm{km~s^{-1}}$
based on the typical nonthermal
velocity dispersions observed in galaxies \citep{braun09}.
We also calculate a 
quantity free from the assumptions on $\Delta v$ by integrating
$\tau_\mathrm{21~cm}$ over the entire line profile:
\begin{eqnarray}
\int {\tau_\mathrm{{21\,cm}}\,\mathrm{d}v} =0.54
\left(\frac{T_\mathrm{s}}{10^3\,\mathrm{K}}\right)^{-1} 
\left(\frac{N_\mathrm{H}}{10^{21}\,\mathrm{cm^{-2}} }
\right)\,\mathrm{km\,s^{-1}}.
\end{eqnarray}


Under a given $N_\mathrm{H}$ and $T_\mathrm{\rm gas}$ at each grid,
we evaluate $f_\mathrm{H_2}=2n_\mathrm{H_2}/n_\mathrm{H}$,
where $n_\mathrm{H}$ and $n_\mathrm{H_2}$ are the number
densities of hydrogen nuclei and H$_2$, respectively.
The formulae are summarized in section 2.1 of \citet{hirashita05}.
We assume the following equilibrium condition:
$R_\mathrm{{dust}}=R_\mathrm{{diss}}$,
where $R_\mathrm{dust}$ and $R_\mathrm{diss}$ are the
rates of
H$_2$ formation on dust grains and H$_2$ dissociation by
UV radiation per unit volume,
respectively. We adopt a dust grain radius of $0.1~\micron$,
a dust material density of
3 g cm$^{-3}$, and the typical UV radiation field in the solar
neighbourhood. The local hydrogen number density is
related to the column density as
$n_\mathrm{H}=N_\mathrm{H}/H$ (the thickness of disk is assumed to
be $H=100$ pc).
Since the dependence of $R_\mathrm{dust}$ on the dust temperature
($T_\mathrm{d}$) is weak, we simply adopt $T_\mathrm{d}=20$ K.
For the dust-to-gas ratio,
we adopt a scaling relation with the metallicity
as $\mathcal{D}=0.01(Z/\mathrm{Z}_{\odot})$.

\subsubsection{Results and Comparison with Observations}\label{subsubsec:result}


\begin{figure}[th]
\begin{center}
\includegraphics[width=0.45\textwidth]{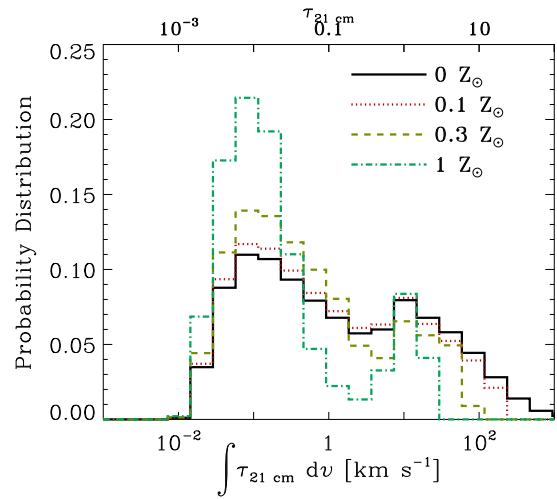}
\end{center}
\caption{Probability distribution function of
$\int \tau_\mathrm{21~cm}\,\mathrm{d}v$ with various
extinctions scaled with the metallicity. The solid,
dotted, dashed, and dot-dashed lines show the
results for $Z = 0$, 0.1, 0.3, and 1 Z$_{\odot}$, respectively.
The upper horizontal axis shows $\tau_\mathrm{21~cm}$
with $\Delta v=10$ km s$^{-1}$.}
\label{fig:prob_tau}
\end{figure}

We show the probability distribution function of $\tau_\mathrm{{21~cm}}$
in Fig.~\ref{fig:prob_tau}. The probability distribution is
calculated
by counting the number of grids in each logarithmic bin of
$\tau_\mathrm{{21~cm}}$ and normalizing the total probability
to unity.
For the case without extinction, we clearly find two peaks around
$\int\tau_\mathrm{21~cm}\,\mathrm{d}v\sim 0.1$ and 10.
The higher peak around  $\tau_\mathrm{{21~cm}} \sim 0.01 $ is 
originating from the warm diffuse medium where column density
is relatively small 
(but larger than $2 \times 10^{20} \mathrm{cm^{-2}} $;
Section \ref{subsubsec:simulation}) with high temperature 
($T_\mathrm{\rm gas}\gtsim 10^3~\mathrm{K}$), 
while the lower peak around $\tau_\mathrm{{21~cm}} \sim 1 $ is due
to cool and dense regions. 
We emphasize that the dominant contribution of the
warm diffuse component is due to the natural consequence of
the compressible hydrodynamical evolution
\citep[see also][]{wada01}, not due to the
stellar feedback, which we neglected. In other words, the
dominant warm diffuse ISM is a natural consequences of
the compressible hydrodynamical evolution of the ISM.

We also show the distribution of
$\tau_\mathrm{{21 cm}}$  with the  dust extinction effect
for $Z = 0.1$, 0.3, and 1 Z$_{\odot}$ in Fig.\ \ref{fig:prob_tau}. It
is obvious that the abundance dramatically decreases at
large $\tau_\mathrm{{21 cm}}$  originating from high column
density regions where
the extinction is large. The peak at  $\tau_\mathrm{{21 cm}} \sim 0.01 $
is relatively enhanced as the extinction bias becomes stronger.

Some features in Fig.\ \ref{fig:prob_tau} match the observational properties of 21 cm 
absorption in DLAs. A large fraction of DLAs are not detected in
21 cm absorption  and typical upper limits of 
$\tau_\mathrm{{21~cm}}$ is roughly $\sim$ 0.02 (or
$\int{\tau_\mathrm{{21~cm}} \,\mathrm{d}v} \sim 0.2~\mathrm{km~s}^{-1}$)
\citep{srianand12,kanekar14}.
The peak of the distribution 
is at $\tau_\mathrm{{21 cm}}\sim$ 0.01, explaining well the non-detection. 
Therefore we conclude that the lack of 21 cm absorption
detection for DLAs is due to the combination of two
effects: a large covering fraction of
the warm diffuse medium and an extinction bias against
large $\tau_\mathrm{21~cm}(\gtsim 1)$.
The possible overestimate of the area of dense regions
(Section \ref{subsubsec:simulation}) strengthens our conclusion
of the significance of the lines of sight with small $\tau_\mathrm{21~cm}$.

\begin{figure}
\begin{center}
\includegraphics[width=0.45\textwidth]{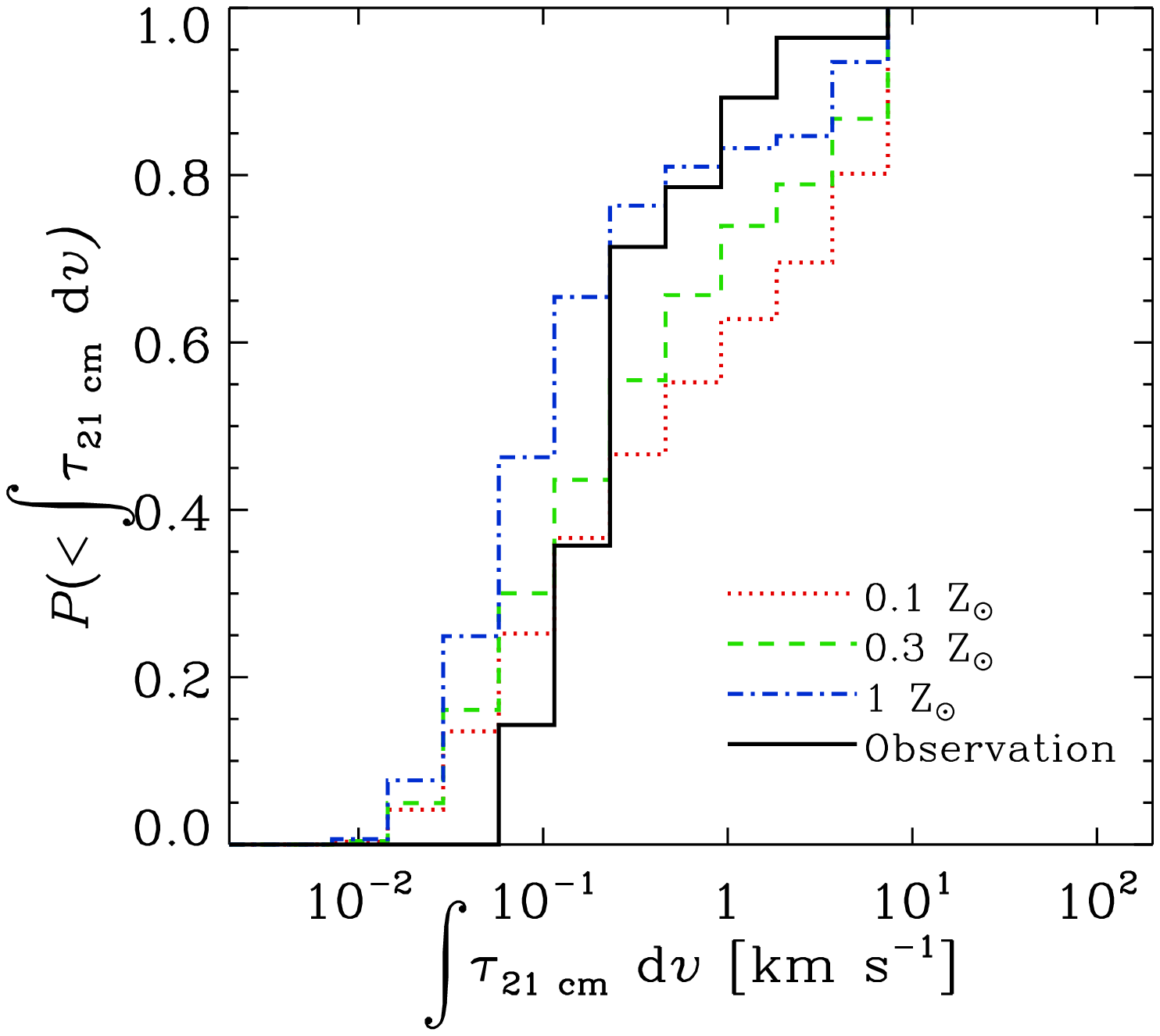}
\end{center}
\caption{Cumulative distribution of $\int \tau_\mathrm{{21 cm}}\,\mathrm{d}v$.
The solid line shows the observational distribution constructed by
the samples in \citet{srianand12} and \citet{kanekar14}.
Note that most of the observational
data are upper limits, which are plotted as detected; thus,
the real observational distribution should be on the left side of
the solid line. The dotted, dashed, and dot-dashed lines show
our theoretical results with $Z=0.1$, 0.3, and 1 Z$_{\odot}$,
respectively. The result for $Z=0$ is almost identical with that
for $Z=0.1$ Z$_{\odot}$.}
\label{fig:comparison} 
\end{figure}

In Fig.\ \ref{fig:comparison}, we show the cumulative distribution
of $\int \tau_\mathrm{{21~cm}}\,\mathrm{d}v$.
The cumulative distribution $P(<\int\tau_\mathrm{21~cm}\,\mathrm{d}v)$
is defined as the probability of having the
integrated $\tau_\mathrm{{21~cm}}$ smaller than a given value.
For the observational data, we refer to \citet{srianand12} and
\citet{kanekar14} with total 48 available DLAs. Since a large fraction of
them only have upper limits of $\tau_\mathrm{21~cm}$, we use the
upper limits for the observed value (solid line in
Fig.\ \ref{fig:comparison}). Thus, the real
distribution should be on the left side of the solid line.
Note that these data are not aimed at complete
sampling so that our attempt of comparison here provides only a first
step for quantitative test. From the figure, we conclude that
at low $\int\tau_\mathrm{{21 cm}}\,\mathrm{d}v$, the theoretical predictions
are consistent with the observation, since they lie on the left side of
the solid line.
This again confirms that the lack of detection of 21 cm absorption for
DLAs is due to the large area covered by small $\tau_\mathrm{21~cm}$.
However, the theoretical prediction overproduces the probability distribution
at large $\int\tau_\mathrm{21 cm}\,\mathrm{d}v$.
The case with the highest metallicity is the most consistent with the
data, since the extinction bias enhances the abundance of DLAs
with small $\tau_\mathrm{21~cm}$.
The discrepancy may be due to the possible overestimate of
the covering fraction of dense regions as mentioned in Section \ref{subsubsec:simulation}.

\subsubsection{Relation between $f_\mathrm{H_2}$ and $\tau_\mathrm{{21~cm}}$}

\begin{figure*}[th]
\begin{center}
		\includegraphics[width=0.44\textwidth]{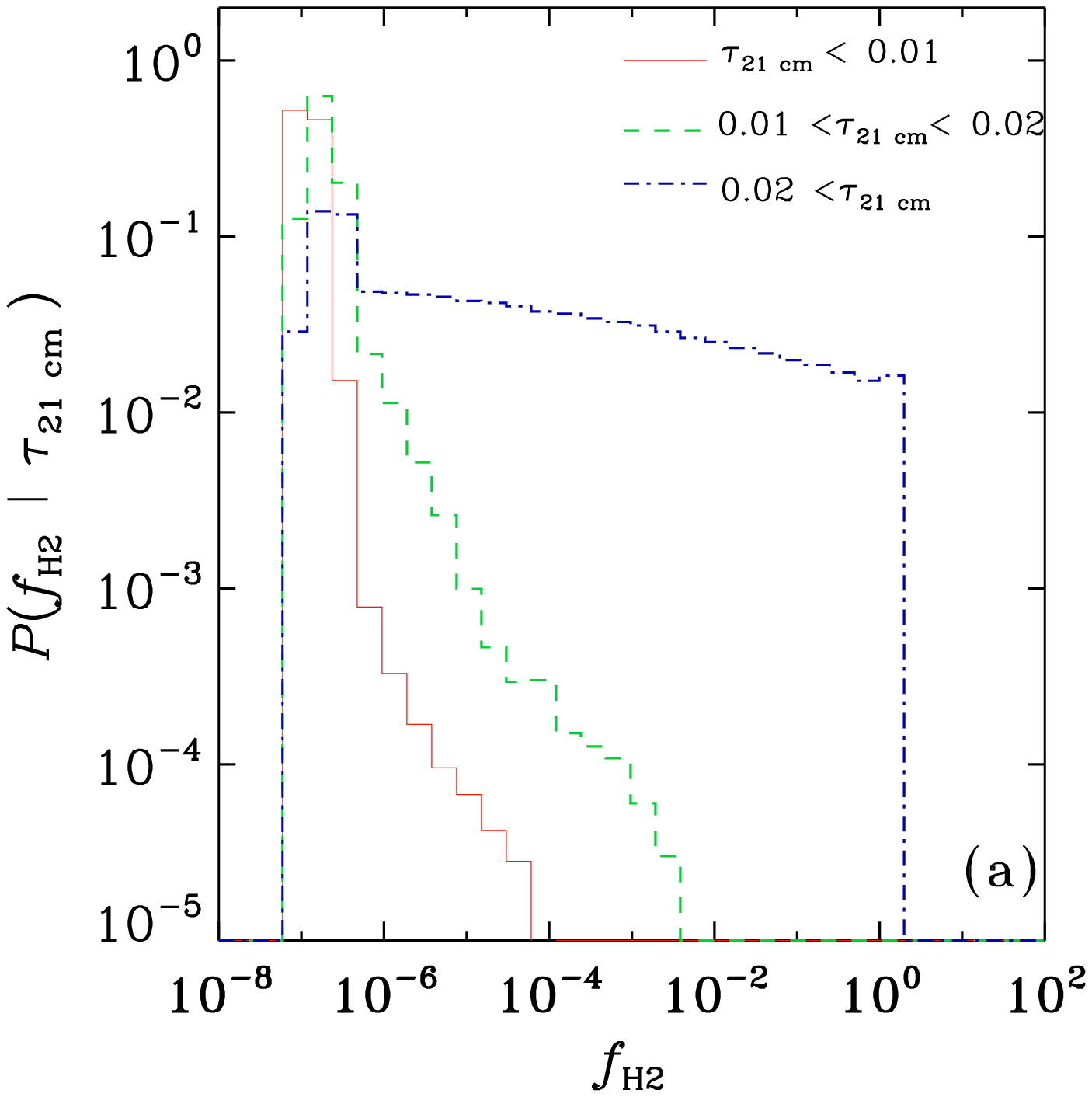}
		\includegraphics[width=0.44\textwidth]{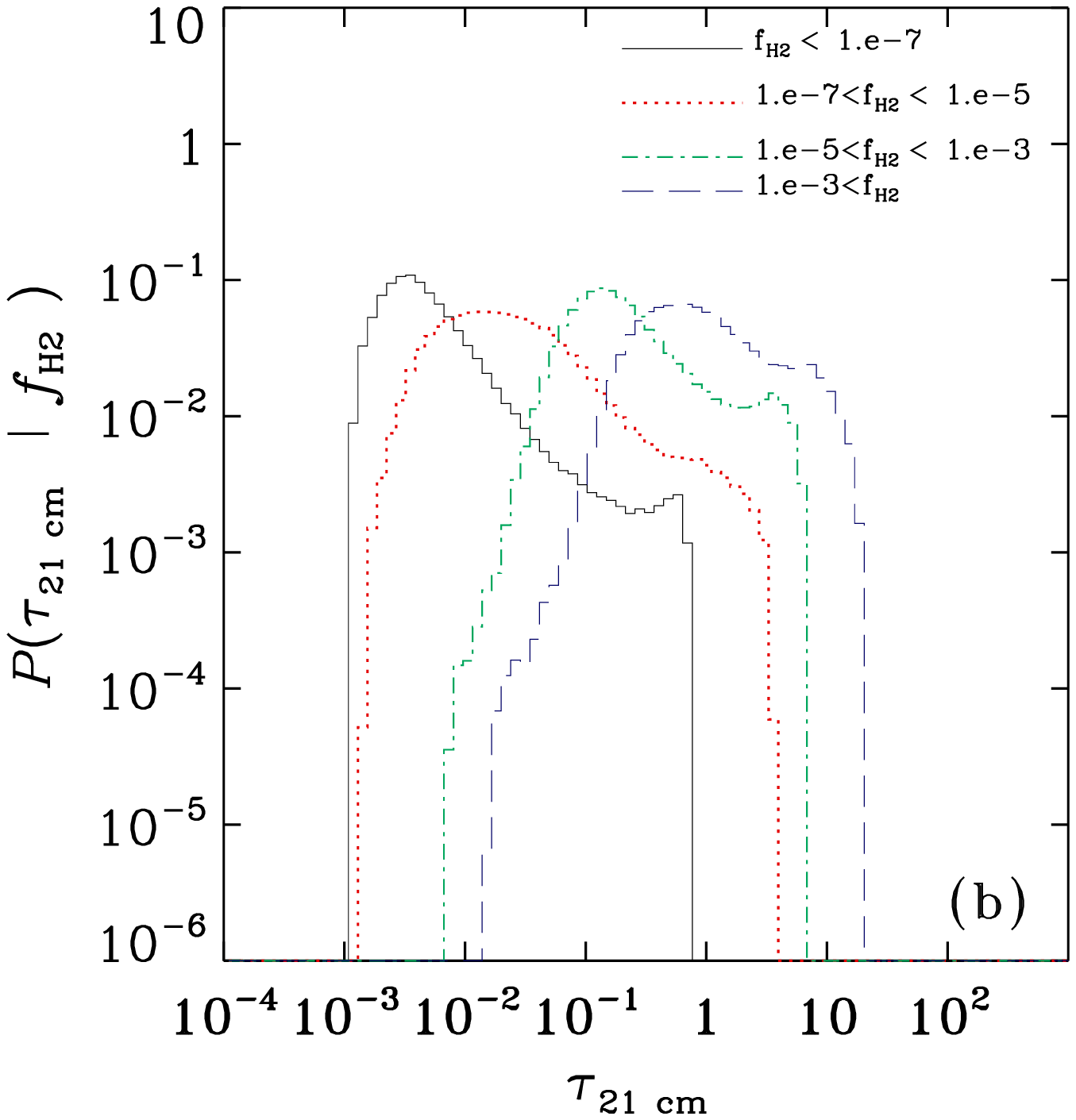}
\end{center}
\caption{(a) Distribution function of $f_\mathrm{H_2}$ for 
$\tau_\mathrm{{21~cm}}<0.01$ (solid line),
$0.01 < \tau_\mathrm{{21~cm}} < 0.02$ (dashed line), and
$\tau_\mathrm{{21~cm}} > 0.02$ (dot-dashed line).
(b) Distribution function of $\tau_\mathrm{21~cm}$ for
$f_\mathrm{H_2}<10^{-7}$ (solid line), $10^{-7}<f_\mathrm{H_2}<10^{-5}$
(dotted line), $10^{-5}<f_\mathrm{H_2}<10^{-3}$ (dot-dashed line),
and $f_\mathrm{H_2}>10^{-3}$.}
\label{fig:fH2_distri}
\end{figure*}

Figure \ref{fig:fH2_distri}a shows the probability distribution function
of $f_\mathrm{H_2}$ for various ranges of $\tau_\mathrm{21~cm}$.
We simply adopt $Z=0.1$ Z$_{\odot}$, but the metallicity does not change
the trend between $\tau_\mathrm{21~cm}$ and $f_\mathrm{H_2}$.
As expected, there is a trend that large $f_\mathrm{H_2}$ appears
as $\tau_\mathrm{21~cm}$ becomes larger.

At $\tau_\mathrm{21~cm}>0.02$ for which the current typical
sensitivity could detect 21 cm absorption, the scatter of
$f_\mathrm{H_2}$ is large as shown in Fig.\ \ref{fig:fH2_distri}a.
Moreover, the peak still lies
at $f_\mathrm{H_2}<10^{-6}$ even for $\tau_\mathrm{21~cm}>0.02$.
On the other hand, a small $\tau_\mathrm{{21~cm}}$ implies a low
$f_\mathrm{H_2}$ ($\ltsim 10^{-6}$). 
Fig.\ \ref{fig:fH2_distri}b shows the probability distribution function
of $\tau_\mathrm{21~cm}$ for various ranges of $f_\mathrm{H_2}$.
We observe that the distribution shifts toward larger
$\tau_\mathrm{21~cm}$ as $f_\mathrm{H_2}$ becomes larger.
This indicates that, if we choose DLAs with
$f_\mathrm{H_2}\gtsim 10^{-3}$, there is a large probability
of detecting 21 cm absorption.
On the other hand, small $f_\mathrm{H_2}<10^{-5}$ does not
necessarily mean that detection of 21 cm absorption is hard,
since there is a significant tail toward $\tau_\mathrm{21~cm}>0.02$
even for $f_\mathrm{H_2}<10^{-5}$.




Our theoretical predictions are summarized as
follows: (i) non-detection of 21 cm absorption
($\tau_\mathrm{21~cm}\ltsim 0.02$) indicates a small H$_2$ fraction
($f_\mathrm{H_2}\ltsim 10^{-6}$). As a corollary of this, a large H$_2$
fraction (especially $f_\mathrm{H_2}\gtsim 10^{-3}$) favours a detectable
21 cm absorption. (ii) Detection of 21 cm ($\tau_\mathrm{21~cm}\gtsim 0.02$)
does not necessarily mean a large H$_2$ fraction.
As a corollary of this, a small H$_2$ fraction ($f_\mathrm{H_2}\ltsim 10^{-5}$)
does not necessarily lead to a non-detection of 21 cm absorption.
These two predictions are broadly supported by the
existing sample in \citet{srianand12} and \citet{kanekar14},
although the number of sample ($\sim 10$) is small.
The exception for (i) is the DLAs
associated with J040718.0$-$441013 ($z_\mathrm{abs}=2.59475$;
$\int\tau_\mathrm{21~cm}\mathrm{d}v<1.61$ km~s$^{-1}$ and
$\log f_\mathrm{H_2}=-2.61$)
and with J053007.9$-$250330 ($z_\mathrm{abs}=2.81115$;
$\int\tau_\mathrm{21~cm}\mathrm{d}v<0.58$ km s$^{-1}$ and
$\log f_\mathrm{H_2}=-2.83$).
However, the upper limits for $\tau_\mathrm{21~cm}$ is still
large (0.16 and 0.058, respectively, if we assume
$\Delta v=10$ km s$^{-1}$). The prediction (ii) is supported
by the current sample. We need to test these predictions
by a larger sample, which would be a critical test for our
theory.


\subsubsection{Requirement for future 21-cm absorption observations
of DLAs}

As shown above, a large part of DLAs are predicted to have
$\tau_\mathrm{21~cm}\ltsim 0.01$ down to $10^{-3}$. Therefore,
in order to detect 21 cm absorption of a major part of DLAs
with 3 $\sigma$, we need to achieve a signal-to-noise ratio of
3000 (i.e., a spectral dynamic range of 35 dB).
Assuming a typical flux of 100 mJy for background QSOs
\citep{srianand12}, a root-mean-square (rms) noise level
of 33 nJy is required. The current rms achieved
by GBT and GMRT is typically a few mJy for an integration
time of $\sim 10$ h \citep[e.g.,][]{srianand12}.
A 100 times larger collecting
area compared with the currently available large radio
telescopes, the same integration time would achieve
an rms of a few $\times 10$ nJy. SKA
would have this kind of large collecting area.

\subsection{Tracing the star formation activities with radio
continuum}\label{sec:model}

We have been contributing to the studies of galaxy evolution
in the radio through modeling of galaxy number count
\citep{takeuchi01} and of spectral energy distribution (SED)
\citep{totani02,hirashita06}. We also plan a survey observation
based on our submillimetre galaxy survey being or to be planned.

\subsubsection{SED models of young starbursts}

As mentioned in Section \ref{subsec:world_conti},
SKA is sensitive enough to conduct wide-area surveys
at $\gtsim 10$ GHz, a relatively high frequency unexplored
for wide-area deep surveys. At such a high frequency,
the relative contribution of free--free emission to
synchrotron emission is larger than at lower frequencies.
In particular, the importance of tracing free--free emission is
pronounced if we target young starbursts.
Indeed, some authors have shown that the radio SEDs
of nearby young starbursts have flat slopes, which
can be interpreted as being dominated by free--free
emission \citep{deeg93,roussel03,hunt05}. However,
the dependence of radio spectral index on age
has not been clarified yet, especially for high-redshift
galaxies. The wide-band coverage of SKA is suitable
to systematically survey the evolution of radio spectral
index as a function of redshift.

\begin{figure}
\begin{center}
\includegraphics[width=0.45\textwidth]{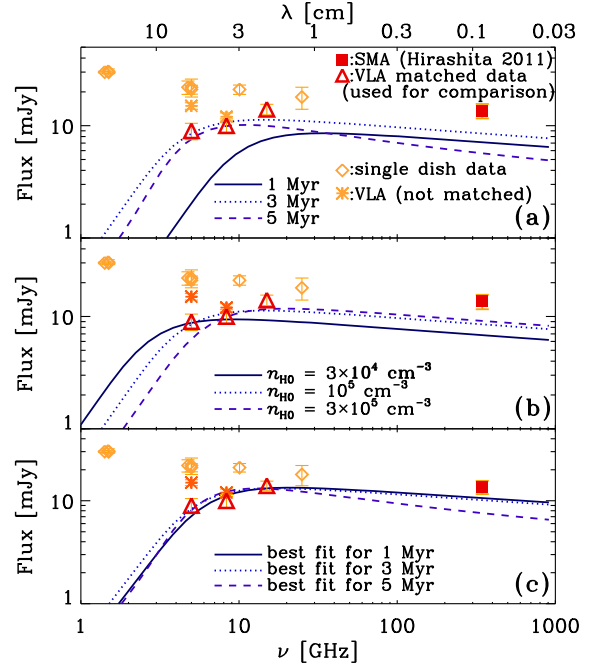}
\end{center}
\caption{Radio SEDs for the central (4$''$) starburst in
II Zw 40 \citep{hirashita11}. (a) Solid, dotted, and dashed
lines show the SEDs calculated by the theoretical model
for ages 1, 3, and 5 Myr, respectively.
The initial hydrogen number density and the total
mass converted into stars are fixed with
$n_\mathrm{H0}=10^5$ cm$^{-3}$ and
$M_0=3\times 10^6~\mathrm{M}_\odot$,
respectively. The filled
square represents our Submillimeter Array (SMA)
measurement at $880~\micron$ for the central
star-forming region \citep{hirashita11}. Diamonds show single-dish
measurements by \citet{jaffe78}, \citet{klein84}, and
\citet{klein91}, while 
asterisks at 6 and 3.6 cm
are the VLA fluxes synthesized by all the $(u,\, v)$
data, which are sensitive to structures up to 10 arcsec and
7 arcsec, respectively
\citep{beck02}: these data are also senstive to emissions
more extended than the central star-forming region.
The triangles at 6, 3.6, and 2 cm are
the matched VLA fluxes obtained by restricting
$(u,\, v)$ data to baselines greater than 20k$\lambda$
(i.e., sensitive to the structures smaller than
4\arcsec) \citep{beck02}. We compare the models with these
data (triangles). (b) Same as Panel (a) but for
the SEDs at
3~Myr with various initial densities. Solid, dotted,
and dashed lines are for $n_\mathrm{H0}=3\times 10^4$,
$10^5$, and $3\times 10^5$ cm$^{-3}$, respectively
($n_\mathrm{H}=1.8\times 10^3$, $3.2\times 10^3$, and
$5.4\times 10^3$\,cm$^{-3}$ at 3 Myr, respectively;
since we consider the expansion of H \textsc{ii} region,
the density becomes lower than the initial value).
(c) Same as Panel (a) but for the
best-fit model parameters
$(M_0\,[\mathrm{M}_\odot ],\, n_\mathrm{H0}\,[\mathrm{cm}^{-3}])=
(1.1\times 10^7,\,2.2\times 10^4)$,
$(3.5\times 10^6,\, 1.2\times 10^5)$, and
$(4.1\times 10^6,\, 1.2\times 10^5)$ at three
given ages, 1, 3, and 5 Myr, for the solid, sotted,
and dashed lines,
respectively. The fitting is applied for the
three triangles (the matched VLA fluxes).
}
\label{fig:iizw40}
\end{figure}

One of the ongoing activities of SKA-Japan is to model
the radio SED based on an empirical approach using
the radio--FIR relation
\citep{takeuchi01} or based on a theoretical approach
treating the evolution of H \textsc{ii} regions and
supernova remnants in a consistent manner with the star
formation history \citep{hirashita06,hirashita11}.
One of the results applied to the central starburst in
a nearby blue compact dwarf galaxy, II Zw 40, is shown in
Figure \ref{fig:iizw40}. We give the total gas mass
$M_0$ and initial hydrogen number density $n_\mathrm{H0}$,
and evaluate the star formation rate as a function of time
as
$\psi (t)=(\epsilon_\mathrm{SF}M_0/t_\mathrm{ff})
\mathrm{e}^{-\epsilon_\mathrm{SF}t/t_\mathrm{ff}}$,
where we adopt the star foramtion efficiency
$\epsilon_\mathrm{SF}=0.1$ and the free-fall time
$t_\mathrm{ff}$ estimated by $n_\mathrm{H,0}$.
The emission measure of the ionized region,
which is necessary to calculate the free--free component,
is estimated consistently with the total number of
ionizing photons produced by massive stars and
the change of hydrogen number density $n_\mathrm{H}$
by the pressure-driven expansion.
The synchrotron component is
also included based on the calculated number of supernova
remnants. Free--free absorption, which is important at
low frequencies, is also considered.

In Figure \ref{fig:iizw40}a, we show the radio SEDs
at $t=1$, 3, and 5\,Myr with
$n_\mathrm{H0}=10^5$ cm$^{-3}$ and
$M_0=3\times 10^6~\mathrm{M}_\odot$. As the age
becomes older, the peak shifts to lower frequencies,
because the free-free optical depth becomes smaller
as the H \textsc{ii} region expands
(the decline of the flux at low frequencies is due to
free--free absorption). At $t=1$ and 3 Myr, the
emission is completely dominated by free--free
emission, and at $t=5~\mathrm{Myr}$, the
synchrotron component begins to contribute to the
emission and the spectrum slope changes.
For comparison, the VLA `matched' data whose
$(u,\, v)$ coverage is restricted to baselines
greater than 20k$\lambda$ (i.e., sensitive to
structures smaller than 4\arcsec) are adopted
(three triangles in Fig.\ \ref{fig:iizw40})
as representative fluxes from the central
star-forming region.

The density strongly affects the frequency at which
the flux peaks because free--free absorption is
sensitive to the density. In Fig.\ \ref{fig:iizw40}b,
we show the SEDs {at $t=3$ Myr} for various
initial densities ($n_\mathrm{H0}$) with
$M_0=3\times 10^6~\mathrm{M}_\odot$. We observe that
the peak position of the SED is indeed
sensitive to the density. The rising
spectrum of the matched data (triangles) is
consistent with free--free absorption.

It is possible to search for the best-fit values of
$M_0$ and $n_\mathrm{H0}$ for each age. The
matched VLA data are adopted (three triangles in
Fig.\ \ref{fig:iizw40}) for the $\chi^2$ fitting,
since our models are applicable to the central
star-forming region. The best-fit solutions are shown
in Fig.\ \ref{fig:iizw40}c.
As seen in Fig.\ \ref{fig:iizw40}c, the
spectral slope at $\nu\gtsim 15~\mathrm{GHz}$ at
$t=5$ Myr is different from that at
$t\leq 3$ Myr because of the
contribution from the synchrotron component.

If the age is about 3 Myr as suggested from the
optical and near-infrared observations
\citep{vanzi08}, 75 per cent of the
$880~\micron$ flux obtained in the SMA observation
\citep{hirashita11}, which is sensitive to the
central starburst, is
explained by free--free emission according to the
best-fit SED at 3 Myr (the flux at $880~\micron$ is
10.2\,mJy in the model, while the observed flux is
13.6\,mJy). The difference ($\simeq 3.4$ mJy) is
likely to be due to the dust and/or diffuse (i.e., not
associated with the
compact H \textsc{ii} region) free--free emission.

The above example indicates that, if we focus on the
young starburst, the radio emission can be dominated
by free--free emission. At the same time, if the
starburst is occuring in a dense and compact region,
which is likely because of the short free--fall time-scale,
emission at $\ltsim 10$ GHz could be significantly affected
by free--free absorption. Therefore, if we would like to
trace a young starburst, it would be desirable to
target frequencies $>10$ GHz.

\subsubsection{Radio survey and combination with
submm data}
\label{subsubsec:survey}

Utilizing the galaxy survey data
already taken by ALMA could make a strong scientific
case, especially because Japan has access to ALMA.
Submm surveys, which will be conducted by ALMA,
are always affected by some bias caused
by dust temperature, as shown below. In contrast,
radio surveys at frequencies that SKA can access have
an advantage of avoiding
a bias caused by dust temperature, if we assume that
the radio--FIR relation always holds.
(In fact, we should keep in mind that there are
other complications at radio wavelengths such as
the variation of spectral
slope, or the relative ratio between free--free and
synchrotron emissions.)

Before considering an SKA survey, we investigate what
kind of preparatory survey can be performed by ALMA.
We focus on 350 GHz ($\lambda =850~\micron$) for the
target frequency, but using a wavelength around 1 mm
does not change the conclusions below.
A shorter wavelength has much lower efficiency because
of a smaller field of view (FoV) and less chance of
suitable atmospheric conditions.
The largest disadvantage of ALMA is a narrow
FoV $\sim 0.047$ arcmin$^2$ at 350 GHz (assuming a
primary beam of 12-m dish), which makes the survey efficiency
extremely low. On the contrary, SKA has a wide FoV.
SKA-MID has a FoV $\sim 37$ arcmin$^2$
(assuming a primary beam of 15-m dish).
Therefore, ALMA needs 790 pointings to
cover a FoV of SKA.

Now let us assume that an on-source observational
time is 1,000 hr for both SKA and ALMA.
Such a long observational time may be appropriate
for future legacy observations of both telescopes.
Then, we can spend
1.3 hours per field by ALMA. Using the ALMA sensitivity
calculator\footnote{https://almascience.eso.org/proposing/sensitivity-calculator}, root mean squeare $\sigma\simeq$ 16 $\mu$Jy
is achieved with 60 antennas at 350 GHz. Therefore, we adopt
80 $\mu$Jy (5$\sigma$) for the detection limit.
On the other hand, we adopt a 5$\sigma$ detection limit of
0.2 $\mu$Jy at 10 GHz for SKA1-MID and 0.02 $\mu$Jy at 10 GHz
for SKA2-MID \citep[e.g.,][]{murphy15}, for an integration
time of 1,000 hr. The detection limit flux of SKA1-MID is
10 times larger.

\begin{figure}
\begin{center}
\includegraphics[width=8cm]{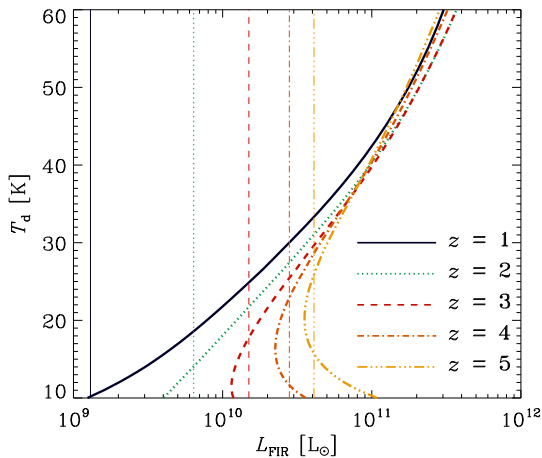}
\end{center}
\caption{Dust temperature ($T_\mathrm{d}$) vs.\
total FIR luminosity ($L_\mathrm{FIR}$) corresponding to
the detection limits for ALMA (350 GHz) and SKA2-MID (10 GHz)
under the condition that a FoV of SKA2-MID
(30 arcmin$^2$) is surveyed in an on-source integration
time of 1,000 hr.
The solid, dotted, dashed, dot-dashed, triple-dot-dashed
lines present the 5$\sigma$ detection limits
galaxies at $z=1$, 2, 3, 4, and 5, respectively.
Objects whose total FIR luminosity is larger than
(i.e., on the right side of) each line are detected.
The thick curves are for ALMA and the thin vertical lines
are for SKA2-MID. The detection limit for SKA1-MID
can be easily estimated by shifting each line for SKA2-MID 
by 10 times towards larger $L_\mathrm{FIR}$.}
\label{fig:flux_Td} 
\end{figure}

To clarify the population expected to be detected
at the ALMA survey described above, we show the relation
between the dust temperature, $T_\mathrm{d}$, and
the total IR luminoisity (total luminosity emitted
by dust, which could be compared to the
8--1000 $\micron$ luminosity), $L_\mathrm{FIR}$,
in Figure \ref{fig:flux_Td},
adopting the 5$\sigma$ detection limit, 80 $\mu$Jy, at
350 GHz.
We adopt the FIR--radio SED model
by \citet{totani02}, who give the SED
under given $T_\mathrm{d}$ and $L_\mathrm{FIR}$.
They also empirically included radio emission
by assuming the observed radio--FIR correlation in
nearby galaxies.\footnote{If we simply use the radio--FIR
relation to estimate the radio luminosity from the total
FIR luminosity, the radio flux should not depend on
the dust temperature. However,
\citet{totani02}'s model, which we adopt, has a weak dependence
of dust temperature through their equation 11.
To avoid this dependence we always adopt a fixed value for
$T_\mathrm{dust}=30$ K in calculating the 10 GHz flux.}
In Figure \ref{fig:flux_Td} (upper panel), we plot the minimum
$L_\mathrm{FIR}$ detected for an object with a
dust temperature $T_\mathrm{d}$; that is, objects with the
FIR luminosity larger than the minimum (or
the right side of the curve) can be detected. From
Figure \ref{fig:flux_Td}, we observe that
the ALMA 350 GHz survey does not have
strong redshift dependence for the dust temperatures
usually observed ($>20$ K), which is due to the
so-called negative $K$-correction, and that the sample is
strongly biased against high $T_\mathrm{d}$.
This kind of bias was already pointed out by
\citet{chapman05}.
On the other hand, SKA2-MID 10 GHz detection limit does not
depend on dust temperature, and is only determined by
FIR luminosity in our model.

In Figure \ref{fig:flux_Td}, we observe that SKA2-MID has
a similar sensitivity to ALMA in terms of
$L_\mathrm{FIR}$ at $z\gtsim 3$ if we assume a normal dust
temperature (20--40 K). Therefore, the combination of
these two surveys are suitable to investigate the
evolution of radio--FIR survey at $z\gtsim 3$. At the same
time, SKA2-MID is also capable of studying a population of low-$L_\mathrm{FIR}$
($L_\mathrm{FIR}\ltsim 10^{11}$ L$_\odot$) galaxies with
high dust temperature $T_\mathrm{dust}\gtsim 30$ K, which are
missed by ALMA surveys. SKA1-MID, which has a ten times worse
sensitivity than SKA2-MID, is suitable to study galaxies
at $z\ltsim 2$ detected by ALMA surveys.

\section{Summary}\label{sec:summary}

Galaxies have been formed from a homogeneous neutral gas in the early 
Universe, and evolved with forming stars from the gas in them. 
Then the key factor of the galaxy formation and evolution is the transition from atomic to
molecular hydrogen. 
This transition is also strongly coupled
with chemical evolution, because dust grains play a critical role in molecular
formation. 
Therefore, a comprehensive understanding of neutral--molecular 
gas transition, star formation and chemical enrichment
is necessary to clarify the galaxy formation and evolution.

The SKA-Japan galaxy evolution subSWG is working on the following 
topics among vast variety of interesting unsolved problems in this field. 
At $z \sim 0$, we try to construct a unified picture
of atomic and molecular hydrogen through nearby galaxies in
terms of metallicity and other various ISM properties. 
In previous studies, the knowledge from the ISM physics was not
efficiently used in extragalactic stuies, mainly because of the limitation
of the angular resolution in radio observations today.
The ultimately high high angular resolution of SKA at cm wavelengths 
will bring us the unprecedented information of the physical state of the 
ISM in galaxies.
By SKA, we will be able to push forward the H\emissiontype{I}--H$_2$ 
transition toward significantly high-$z$ and discuss its evolution. 

Up to intermediate redshifts $z \sim 1$, we explore scaling relations
including gas and star formation properties, like the main
sequence and the Kennicutt--Schmidt law of star forming galaxies. 
{}To connect the global studies with spatially-resolved investigations,
such relations will be plausibly a viable way. 
Especially in relation to the H\emissiontype{I}--H$_2$ topic, the 
Kennicutt--Schmidt law provides a unique opportunity, since 
this relation holds  both for resolved (local) and global galaxy 
properties. 
If we can trace the evolution (or constancy) of these 
scaling relations, we can constrain the star formation scenarios
in galaxy evolution. 
It is particularly important since it is the way how to connect the ISM physics 
and extragalactic physics and organize them on the common basis. 
Only SKA can achieve such observations.

At high redshift, it would be very difficult to detect the H\emissiontype{I} emission 
of galaxies, even with SKA2. 
Instead, the absorption lines of HI 21-cm line will be a very
promising observable to explore the properties of gas in galaxies.
We are interested in the physical processes going on during the galaxy formation. 
One of the ideal proves for this is a high column density system like DLAs. 
Though DLAs are alreacy extensively studied in the field of galaxy formation and
cosmology, one possible bias was pointed out.
If a DLA contains even a tiny amount of dust, it would bias the detection
probability of background quasars. 
This plausible bias can be compensated by selecting radio-loud quasars and
explore the 21-cm absorption line system. 
Since the cross section of such system is small, we can examine their 
physical properties even if the column density is extremely high. 

Apart from H\emissiontype{I}, we have yet another great probe for the
star formation history in the whole range of the cosmic time. 
The radio continuum at $\nu \gtsim 10$~GHz consists of thermal free-free
and synchrotron emission. 
Both of these radiation processes are related to the star formation
activity in galaxies, and work as a good tracer of the SFR.
Since the feasibility of the radio continuum by SKA is much higher 
than that of H\emissiontype{I} 21-cm line, the radio continuum is  
very suitable for the exploration of the cosmic SFR from the very 
early phase of galaxy formation. 
 
By these studies, we will surely witness a real revolution in
the studies of galaxies by SKA.

\begin{ack}
We are grateful to K.\ Kohno for useful comments.
TTT has been supported by the Grant-in-Aid for the Scientific Research 
Fund (23340046) commissioned by the Ministry of Education, Culture, 
Sports, Science and Technology (MEXT) of Japan.
HH thanks the support from the Ministry of Science and Technology
(MoST) grant 102-2119-M-001-006-MY3.
\end{ack}

\bibliographystyle{aa}
\bibliography{paper}

\begin{thebibliography}{99}


\bibitem[Kalberla \& Kerp(2009)]{kk09}
	Kalberla, P. M. W., \& Kerp, J., 2009, ARA\&A, 47, 27
\bibitem[Kaneko et al.(2013)]{kaneko13}
	Kaneko, H., et al., 2013, PASJ, 65, 20	
\bibitem[Nakanishi \& Sofue(2016)]{ns15}
	Nakanishi, H. \& Sofue, Y., 2016, PASJ, 68, 5
\bibitem[Nakanishi et al.(2006)]{nakanishi06}
	Nakanishi, H. et al., 2006, ApJ, 651, 804
\bibitem[Sofue \& Rubin(2001)]{sofue01}
	Sofue, Y. \& Rubin, V., 2001, ARA\&A, 39, 137
\bibitem[Tosaki et al.(2011)]{tosaki11}
	Tosaki, T. et al., 2011, PASJ, 63, 1171


\bibitem[\protect\citeauthoryear{Akerman et al.}{2005}]{akerman05}
    Akerman, C. J., Ellison, S. L., Pettini, M., \& Steidel, C. C.
    2005, A\&A, 440, 499
\bibitem[Allen et al.(2003)]{2003MNRAS.342..287A} 
	Allen, S.~W. et al.\ 2003, MNRAS, 342, 2887 
\bibitem[\protect\citeauthoryear{Anders \& Grevesse}{1989}]{anders89}
    Anders E., Grevesse N., 1989, Geochim.\ Cosmochim.\ Acta, 53, 197
\bibitem[\protect\citeauthoryear{Asano et al.}{2013}]{asano13}
    Asano, R. S., Takeuchi, T. T., Hirashita, H., Nozawa, T. 2013,
    MNRAS, 432, 637
\bibitem[Asano et al.(2014)]{asano2014}
	Asano, R. S., Takeuchi, T. T., Hirashita, H., \& Nozawa, T. 2014,
	MNRAS, 440, 134
\bibitem[Babul \& Rees(1992)]{1992MNRAS.255..346B}
	Babul A. \& Rees M. J., 1992, MNRAS, 255, 346
\bibitem[Bacon et al.(2001)]{2001MNRAS.326...23B}
	Bacon R. et al., 2001, MNRAS, 326, 23
\bibitem[Barnes et al.(2001)]{2001MNRAS.322..486B}
	Barnes J. E. et al., 2001, MNRAS, 322, 486
\bibitem[\protect\citeauthoryear{Beck et al.}{2002}]{beck02}
    Beck, S. C., Turner, J. L., Langland-Shula, L. E., Meier, D. S., Crosthwaite, L. P., \&
    Gorjian, V. 2002, AJ, 124, 2516
\bibitem[Begum et al.(2008)]{2008MNRAS.386.1667B}
	Begum A. et al., 2008, MNRAS, 386, 1667
\bibitem[Benson et al.(2002)]{2002MNRAS.333..156B}
	Benson A. J., et al., 2002, MNRAS, 333, 156
\bibitem[Bell \& de Jong(2001)]{bdj01}	
	Bell, E. F. \& de Jong, R. S., 2001, ApJ, 550, 212
\bibitem[\protect\citeauthoryear{Bekki}{2015}]{bekki15}
    Bekki, K. 2015, MNRAS, 449, 1625
\bibitem[Bigiel et al.(2008)]{bigiel08}	
	Bigiel, F. et al., 2008, AJ, 136, 2846
\bibitem[Bigiel et al.(2011)]{bigiel11}	
	Bigiel, F., et al., 2011, ApJ, 730, 13
\bibitem[Bigiel \& Blitz(2012)]{bb12}	
	Bigiel, F. \& Blitz, L., 2012, ApJ, 756, 183
\bibitem[\protect\citeauthoryear{Black, Chaffee, \& Foltz}{1987}]{black87}
    Black J. H., Chaffee F. H. Jr., Foltz C. B., 1987, ApJ, 317, 442
\bibitem[Blitz \& Rosolowsky(2004)]{br04}	
	Blitz, L. \& Rosolowsky, E., 2004, ApJ, 612, 29
\bibitem[Boselli et al.(2002)]{boselli02}
	Boselli, A. et al., 2002, A\&A, 564, 66
\bibitem[Bosma et al.(1981)]{bosma81}	
	Bosma, A., 1981, AJ, 86, 1791
\bibitem[\protect\citeauthoryear{Briggs, Brinks, \& Wolfe}{1997}]{briggs97}
    Briggs F. H., Brinks E., Wolfe A. M., 1997, AJ, 113, 467
\bibitem[\protect\citeauthoryear{Braun}{2012}]{braun12}
    Braun R., 2012, ApJ, 749, 87
\bibitem[\protect\citeauthoryear{Braun et al.}{2009}]{braun09}
    Braun R., Thilker D. A., Walterbos R. A. M., Corbelli E. 2009, ApJ, 695, 937
\bibitem[Buat et al.(2005)]{buat2005} 
	Buat, V., Iglesias-P{\'a}ramo, J., Seibert, M., et al.\ 2005, ApJ, 619, L51 
\bibitem[Buat et al.(2007)]{buat2007} 
	Buat, V., Takeuchi, T.~T., Iglesias-P{\'a}ramo, J., et al.\ 2007, ApJS, 173, 404 
\bibitem[Butcher \& Oemler(1978)]{1978ApJ...219...18B} 
	Butcher, H. and Oemler, Jr., A.\ 1978, ApJ, 219, 18 
\bibitem[Butcher et al.(1983)]{1983ApJS...52..183B} 
	Butcher, H., Wells, D.~C. and Oemler, Jr., A.\ 1983, ApJS, 52, 183 
\bibitem[Cannon et al.(2011)]{2011ApJ...739L..22C}
	Cannon J. M. et al., 2011, ApJ, 739, 22
\bibitem[Cantinella et al.(2010)]{2010MNRAS.403..683C}
	Cantinella B. et al., 2010, MNRAS, 403, 683
\bibitem[Cantinella et al.(2012a)]{2012A&A...544A..65C}
	Catinella B. et al., 2012b, A\&A, 544, 65
\bibitem[Cantinella et al.(2012b)]{2012MNRAS.420.1959C}
	Catinella B. et al., 2012a, MNRAS, 420, 1959
\bibitem[Cantinella et al.(2013)]{2013MNRAS.436...34C}
	Catinella B. et al., 2013, MNRAS, 436, 34
\bibitem[Cappellari et al.(2011)]{2011MNRAS.416.1680C}
	Cappellari M. et al., 2011, MNRAS, 416, 1680
\bibitem[\protect\citeauthoryear{Carilli et al.}{1996}]{carilli96}
    Carilli C. L., Lane W., de Bruyn A. G., Braun R., Miley G. K.,
    1996, AJ, 111, 1830
\bibitem[\protect\citeauthoryear{Casey et al.}{2015}]{casey15}
    Casey, C. M., et al.\ 2015, Next Generation Very Large Array Memos Series,
    No.\ 8 (arXiv:1510.06411)
\bibitem[\protect\citeauthoryear{Chapman et al.}{2005}]{chapman05}
    Chapman, S. C., Blain, A. W., Smail, I., \&
    Ivison, R. J. 2005, ApJ, 622, 772
\bibitem[\protect\citeauthoryear{Condon}{1992}]{condon92}
    Condon, J. J. 1992, ARA\&A, 30, 575
\bibitem[\protect\citeauthoryear{Curran et al.}{2010}]{curran10}
    Curran S. J., Tzanavaris P., Darling J. K., Whiting M. T.,
    Webb J. K., Bignell C., Athreya R., Murphy M. T., 2010, MNRAS, 402, 35
\bibitem[Dave et al.(2011)]{dave11}	
	Dave R., Oppenheimer B. D. \& Finlator K., 2011, MNRAS, 415, 11
\bibitem[de Blok et al.(2003)]{2003MNRAS.340..657D}
	de Blok W. J. G. et al., 2003, MNRAS, 340, 657
\bibitem[de Blok et al.(2005)]{2005ApJ...634..227D}
	de Blok W. J. G., 2005, ApJ, 634, 227
\bibitem[\protect\citeauthoryear{de Bruyn, O'Dea, \& Baum}{1996}]{debruyn96}
    de Bruyn A. G., O'Dea C. P., Baum S. A., 1996, A\&A, 305, 450
\bibitem[\protect\citeauthoryear{Deeg et al.}{1993}]{deeg93}
    Deeg, H.-J., Brinks, E., Duric, N., Klein, U., \& Skillman, E.
    1993, ApJ, 410, 626
\bibitem[Dekel et al.(2009a)]{dekel09a}	
	Dekel A. et al., 2009a, Nature, 457, 451
\bibitem[Dekel et al.(2009b)]{dekel09b}	
	Dekel A., Sari R. \& Ceverino D., 2009b, ApJ, 703, 785
\bibitem[\protect\citeauthoryear{Draine \& Bertoldi}{1996}]{draine96}
    Draine B. T., Bertoldi F., 1996, ApJ, 468, 269
\bibitem[\protect\citeauthoryear{Ellison, Hall, \& Lira}{Ellison et al.}{2005}]{ellison05}
    Ellison, S. L., Hall, P. B., \& Lira, P. 2005, AJ, 130, 1345
\bibitem[Elmegreen(1993)]{elmegreen93}	
	Elmegreen, B. G., 1993, ApJ, 411, 170
\bibitem[\protect\citeauthoryear{Fall \& Pei}{1993}]{fall93}
    Fall, S. M., \& Pei, Y. C. 1993, ApJ, 402, 479
\bibitem[Fern{\'a}ndez et al.(2013)]{2013ApJ...770L..29F}
	Fern{\'a}ndez, X. et al., 2013, ApJ, 770, 29
\bibitem[Ferrara \& Tolstoy(2000)]{2000MNRAS.313..291F}
	Ferrara A. \& Tolstoy E., 2000, MNRAS, 313, 291
\bibitem[\protect\citeauthoryear{Field}{1958}]{field58}
    Field G. B., 1958, Proc.\ I.R.E., 46, 240
\bibitem[Fumagalli et al.(2010)]{fumagalli10}	
	Fumagalli, M., Krumholz, M. R. \& Hunt, L. K., 2010, ApJ, 722, 919
\bibitem[Fukui et al.(2015)]{fukui15}	
	Fukui, Y. et al., 2015, ApJ, 798, 6
\bibitem[\protect\citeauthoryear{Furlanetto, Oh, \& Briggs}{Furlanetto et al.}{2006}]{furlanetto06}
    Furlanetto S. R., Oh S. P., Briggs F. H., 2006, Phys.\ Rep., 433, 181
\bibitem[\protect\citeauthoryear{Fynbo et al.}{2011}]{fynbo11}
    Fynbo J. P. U., et al., 2011, MNRAS, 413, 2481
\bibitem[\protect\citeauthoryear{Garrett}{2002}]{garrett02}
    Garrett, M. A. 2002, A\&A, 384, L19
\bibitem[Genzel et al.(2012)]{genzel2012} 
	Genzel, R., Tacconi, L.~J., Combes, F., et al.\ 2012, ApJ, 746, 69 
\bibitem[Giovanelli \& Haynes(1983)]{1983AJ.....88..881G}
	Giovanelli R. \& Haynes M. P., 1983, AJ, 88, 881
\bibitem[Giovanelli et al.(2005)]{2005AJ....130.2598G}
    	Giovanelli R. et al., 2005, AJ, 130, 2598
\bibitem[Glover \& Mac Low(2007)]{gml07}
	Glover, S. C. O. \& Mac Low, M., 2007, ApJ, 659, 1317
\bibitem[Glover \& Clark(2012)]{gk12}
	Glover, S. C. O. \& Clark, P. C., 2012, MNRAS, 421, 9
\bibitem[Gnedin et al.(2009)]{gnedin09}	
	Gnedin, N. Y., Tassis, K. \& Kravtsov, A. V., 2009, ApJ, 697, 55
\bibitem[Gouguenheim et al.(1969)]{1969A&A.....3..281G}
	Gouguenheim L., 1969, A\&A, 3, 281
\bibitem[Gould \& Salpeter(1963)]{gs63}
	Gould, R. J. \& Salpeter, E. E., 1963, ApJ, 138, 393
\bibitem[Grebel et al.(2003)]{2003AJ....125.1926G}
	Grebel E. K. et al., 2003, AJ, 125, 1926
\bibitem[Grossi et al.(2009)]{2009A&A...498..407G}
	Grossi M. et al., 2009, A\&A, 498, 407
\bibitem[\protect\citeauthoryear{Gruppioni et al.}{2003}]{gruppioni03}
    Gruppioni, C., Pozzi, F., Zamorani, G., Ciliegi, P., Lari, C.,
    Calabrese, E., La Franca, F., Matute, I. 2003, MNRAS, 341, L1
\bibitem[Hibbard et al.(2001)]{2001ASPC..240..657H}
	Hibbard J. E. et al., 2001, ASPC, 240, 657
\bibitem[\protect\citeauthoryear{Hirashita et al.}{2003}]{hirashita03}
    Hirashita H., Ferrara A., Wada K., Richter P., 2003, MNRAS, 341, L18 (H03)
\bibitem[\protect\citeauthoryear{Hirashita \& Ferrara}{2005}]{hirashita05}
    Hirashita H., Ferrara A., 2005, MNRAS, 356, 1529
\bibitem[\protect\citeauthoryear{Hirashita \& Hunt}{2006}]{hirashita06}
    Hirashita H., \& Hunt, L. K. 2006, A\&A, 460, 67
\bibitem[\protect\citeauthoryear{Hirashita}{2011}]{hirashita11}
    Hirashita, H. 2011, MNRAS, 418, 828
\bibitem[\protect\citeauthoryear{Hirashita et al.}{2015}]{hirashita15}
    Hirashita, H., et al.\ 2015, PASJ, submitted
\bibitem[Hoeft et al.(2006)]{2006MNRAS.371..401H}
	Hoeft M. et al., 2006, MNRAS, 371, 401
\bibitem[\protect\citeauthoryear{Hollenbach \& McKee}{1979}]{hollenbach79}
    Hollenbach D. J., McKee C. F., 1979, ApJS, 41, 555
\bibitem[Hollenbach \& Salpeter(1971)]{hs71}
	Hollenbach, D. \& Salpeter, E. E., 1971, ApJ, 163, 155
\bibitem[Honma et al.(1995)]{honma95}	
	Honma, M., Sofue, Y. \& Arimoto, N., 1995, A\&A, 304, 1
\bibitem[\protect\citeauthoryear{Hunt, Dyer, \& Thuan}{Hunt et al.}{2005}]{hunt05}
    Hunt, L. K., Dyer, K. K., \& Thuan, T. X. 2005, A\&A, 436, 837
\bibitem[Hunter et al.(2012)]{2012AJ....144..134H}
	Hunter D. A. et al., 2012, AJ, 144, 134
\bibitem[\protect\citeauthoryear{Ibar et al.}{2008}]{ibar08}
    Ibar, E., et al.\ 2008, MNRAS, 386, 953
\bibitem[\protect\citeauthoryear{Inoue, Hirashita, \& Kamaya}{Inoue et al.}{2000}]{inoue00}
    Inoue, A. K., Hirashita, H., \& Kamaya, H. 2000, PASJ, 52, 539
\bibitem[\protect\citeauthoryear{Inoue, Omukai, \& Ciardi}{Inoue et al.}{2007}]{inoue07}
    Inoue, S., Omukai, K., \& Ciardi, B. 2007, MNRAS, 280, 1715
\bibitem[\protect\citeauthoryear{Jaffe, Perola, \& Tarenghi}{Jaffe et al.}{1978}]{jaffe78}
    Jaffe, W. J., Perola, G. C., \& Tarenghi, M. 1978, ApJ, 224, 808
\bibitem[Jaff{\'e} et al.(2012)]{2012ApJ...756L..28J} 
	Jaff{\'e}, Y.~L. et al.\ 2012, ApJ, 756, L28 
\bibitem[Jaff{\'e} et al.(2013)]{2013MNRAS.431.2111J}
	Jaff{\'e}, Y. L. et al., 2013, MNRAS, 431, 2111
\bibitem[\protect\citeauthoryear{Jarvis et al.}{2015}]{jarvis15}
    Jarvis, M. J., et al.\ 2015, in Proceednigs of
    Advancing Astrophysics with the Square Kilometre Array (AASKA14), 68
\bibitem[Johnsen \& Guberman(2010)]{jg10}
	Johnsen, R. \& Guberman, S. L., 2010, 
	Elsevier Inc., "Advances In Atomic, Molecular, and Optical Physics", 59, 75
\bibitem[\protect\citeauthoryear{Kanekar et al.}{2014}]{kanekar14}
    Kanekar N., et al., 2014, MNRAS, 438, 2131
\bibitem[\protect\citeauthoryear{Kanekar et al.}{2013}]{kanekar13}
    Kanekar N., Ellison S. L., Momjian E., York B. A., Pettini M.
    2013, MNRAS, 428, 532
\bibitem[\protect\citeauthoryear{Kanekar et al.}{2006}]{kanekar06}
    Kanekar N., Subrahmanyan R., Ellison S. L., Lane W. M.,
    Chengalur J. N., 2006, MNRAS, 370, L46
\bibitem[\protect\citeauthoryear{Kennicutt}{1998}]{kennicutt98}
    Kennicutt, R. C., Jr.\ 1998, ARA\&A, 36, 189
\bibitem[Kennicutt \& Evans(2012)]{kennicutt2012} 
	Kennicutt, R.~C., \& Evans, N.~J.\ 2012, ARA\&A, 50, 531 
\bibitem[Kennicutt et al.(2007)]{kennicutt07}
	Kennicutt, R. C., et al., 2007, ApJ, 671, 333	
\bibitem[\protect\citeauthoryear{Klein, Weiland, \& Brinks}{Klein et al.}{1991}]{klein91}
    Klein, U., Weiland, H., \& Brinks, E. 1991, A\&A, 246, 323
\bibitem[\protect\citeauthoryear{Klein, Wielebinski, \& Thuan}{Klein et al.}{1984}]{klein84}
    Klein, U., Wielebinski, R., \& Thuan, T. X. 1984, A\&A, 141, 241
\bibitem[Knapp et al.(1985)]{1985AJ.....90..454K}
	Knapp G. R. et al., 1985, AJ, 90, 454
\bibitem[Krumholz et al.(2008)]{krumholz08}
	Krumholz, M. R., McKee, C. F. \& Tumlinson, J., 2008, ApJ, 689, 865
\bibitem[Krumholz et al.(2009)]{krumholz09}	
	Krumholz, M. R., McKee, C. F. \& Tumlinson, J., 2009, ApJ, 693, 216
\bibitem[Krumholz(2012)]{krumholz12}
	Krumholz M. R., 2012, ApJ, 759, 9
\bibitem[Krumholz(2013)]{krumholz13}
	Krumholz M. R., 2013, IAUS, 292, 227
\bibitem[Kuntschner et al.(2010)]{2010MNRAS.408...97K}
	Kuntschner H. et al., 2010, MNRAS, 408, 97
\bibitem[Kwan(1977)]{kwan77}
	Kwan. J., 1977, ApJ, 216, 713

\bibitem[\protect\citeauthoryear{Ledoux, Petitjean, \& Srianand}{Ledoux et al.}{2003}]{ledoux03}
    Ledoux C., Petitjean P., Srianand R., 2003, MNRAS, 346, 209
\bibitem[Lemonias et al.(2013)]{2013ApJ...776...74L}
	Lemonias J. J. et al., 2013, ApJ, 776, 74
\bibitem[Leroy et al.(2008)]{leroy08}	
	Leroy, A. K. et al., 2008, AJ, 136, 2782
\bibitem[\protect\citeauthoryear{Levshakov et al.}{2000}]{levshakov00}
    Levshakov S. A., Molaro P., Centuri\'{o}n M., D'Odorico S.,
    Bonifacio P., Vladilo G., 2000, A\&A, 361, 803
\bibitem[Lewis et al.(2002)]{2002MNRAS.334..673L}
	Lewis I. et al., 2002, MNRAS, 334, 673
\bibitem[Liszt(2002)]{liszt02}
	Liszt, H., 2002, A\&A, 389, 393
\bibitem[London et al.(1977)]{london77}	
	London. R., McCray. R. \& Chu, S.-I., 1977, ApJ, 217, 442
\bibitem[Mac Low \& Ferrara(1999)]{1999ApJ...513..142M}
	Mac Low M. \& Ferrara A., 1999, ApJ, 513, 142
\bibitem[Mac Low \& Glover(2012)]{mlg12}
	Mac Low, M. \& Glover, S. C. O., 2012, ApJ, 746, 135
\bibitem[Magnelli et al.(2012)]{magnelli2012} 
	Magnelli, B., Saintonge, A., Lutz, D., et al.\ 2012, A\&A, 548, AA22 
\bibitem[Magnelli et al.(2014)]{magnelli2014} 
	Magnelli, B., Lutz, D., Saintonge, A., et al.\ 2014, A\&A, 561, AA86 
\bibitem[Magorrian et al.(1998)]{magorrian1998}
	Magorrian, J., Tremaine, S., Richstone, D., et al.\ 1998, AJ, 115, 2285
\bibitem[Martin et al.(2010)]{2010ApJ...723.1359M}
	Martin A. M. et al., 2010, ApJ, 723, 1359
\bibitem[McGaugh et al.(2000)]{2000ApJ...533L..99M}
	McGaugh S. S. et al., 2000, ApJ, 533, 99
\bibitem[McGaugh et al.(2005)]{2005ApJ...632..859M}
	McGaugh S. S., 2005, ApJ, 632, 859
\bibitem[McGaugh(2012)]{mcgaugh12}	
	McGaugh, S. S., 2012, AJ, 143, 40
\bibitem[McKee \& Krumholz(2010)]{mk10}	
	McKee, C. F. \& Krumholz, M. R., 2010, ApJ, 709, 308
\bibitem[\protect\citeauthoryear{Micha{\l}owski, Watson, \& Hjorth}{Micha{\l}owski et al.}{2010}]{michalowski10}
    Micha{\l}owski, M. J., Watson, D., \& Hjorth, J. 2010, ApJ, 712, 942
\bibitem[Momose et al.(2013)]{momose2013} 
	Momose, R., Koda, J., Kennicutt, R.~C., Jr., et al.\ 2013, ApJ, 772, LL13  
\bibitem[Moran et al.(2012)]{2012ApJ...745...66M}
	Moran S.~M. et al., 2012, ApJ, 745, 66
\bibitem[Morganti et al.(2006)]{2006MNRAS.371..157M}
	Morganti R. et al., 2006, MNRAS, 371, 157
\bibitem[\protect\citeauthoryear{Morganti, Sadler, \& Curran}{Morganti et al.}{2015}]{morganti15}
    Morganti, R., Sadler, E. M., \& Curran, S. 2015, in Proceednigs of
    Advancing Astrophysics with the Square Kilometre Array (AASKA14), 134
\bibitem[Mori et al.(2002)]{mori02}	
	Mori M., Ferrara A. \& Madau P., 2002, ApJ, 571, 40
\bibitem[Mori \& Umemura(2006)]{mori06}	
	Mori M. \& Umemura M., 2006, Nature, 440, 644
\bibitem[\protect\citeauthoryear{Morrison et al.}{2010}]{morrison10}
    Morrison, G. E., Owen, F. N., Dickinson, M., Ivison, R. J., \&
    Ibar, E. 2010, ApJS, 188, 178
\bibitem[\protect\citeauthoryear{Murphy}{2009}]{murphy09}
    Murphy, E. J. 2009, ApJ, 706, 482
\bibitem[\protect\citeauthoryear{Murphy et al.}{2015}]{murphy15}
    Murphy, E. J., et al.\ 2015, in Proceednigs of
    Advancing Astrophysics with the Square Kilometre Array (AASKA14), 85
\bibitem[Navarro et al.(2004)]{2004MNRAS.349.1039N}
	Navarro J. F. et al., 2004, MNRAS, 349, 1039
\bibitem[Neufeld \& Spaans(1996)]{ns96}
	Neufeld, D. A. \& Spaans, M., 1996, ApJ, 473, 894
\bibitem[\protect\citeauthoryear{Okamoto}{2013}]{okamoto13}
    Okamoto, T. 2013, MNRAS, 428, 718
\bibitem[\protect\citeauthoryear{Omukai}{2000}]{omukai00}
    Omukai K., 2000, ApJ, 534, 809
\bibitem[Omukai et al.(2012)]{omukai12}	
	Omukai, K., et al., 2010, ApJ, 722, 1793
\bibitem[Oosterloo et al.(2010)]{2010MNRAS.409..500O}
	Oosterloo T. et al., 2010, MNRAS, 409, 500
\bibitem[Ott et al.(2012)]{2012AJ....144..123O}
	Ott J. et al., 2012, AJ, 144, 123
\bibitem[Pelupessy et al.(2006)]{pelupessy06}	
	Pelupessy, F. I., Papadopoulos, P. P. \& van der Werf, P., 2006, ApJ, 645, 1024
\bibitem[\protect\citeauthoryear{P\'{e}roux et al.}{2003}]{peroux03}
    P\'{e}roux C, McMahon R. G., Storrie-Lombondi L. L., Irwin M. J.,
    2003, MNRAS, 346, 1003
\bibitem[\protect\citeauthoryear{Petitjean et al.}{2006}]{petitjean06}
    Petitjean P., Ledoux C., Noterdaeme P., Srianand R., 2006, A\&A, 456, L9
\bibitem[\protect\citeauthoryear{Petitjean, Srianand, \& Ledoux}{Petitjean et al.}{2000}]{petitjean00}
    Petitjean P., Srianand R., Ledoux C., 2000, A\&A, 364, L26
\bibitem[Pisano(2014)]{pisano14}	
	Pisano, D. J., 2014, AJ, 147, 48
\bibitem[Prandoni \& Seymour(2015)]{prandoni15}
    Prandoni, I., \& Seymour, N. 2015, in Proceednigs of
    Advancing Astrophysics with the Square Kilometre Array (AASKA14),  67
\bibitem[\protect\citeauthoryear{Prochaska \& Wolfe}{2000}]{prochaska00}
    Prochaska J. X., Wolfe A. M., 2002, A\&A, 566, 68
\bibitem[\protect\citeauthoryear{Razoumov et al.}{2006}]{razoumov06}
    Razoumov, A. O., Norman, M., Prochaska, J. X., \& Wolfe, A. M. 2006, ApJ, 645, 5
\bibitem[Rees(1986)]{1986MNRAS.218P..25R}
	Rees M. J., 1986, MNRAS, 218, 25
\bibitem[Richings et al.(2014)]{richings14}	
	 Richings, A. J., Schaye, J. \& Oppenheimer, B. D., 2014, MNRAS, 442, 2780
\bibitem[Roberts \& Haynes(1994)]{roberts1994} 
	Roberts, M.~S., \& Haynes, M.~P.\ 1994, ARA\&A, 32, 115 
\bibitem[\protect\citeauthoryear{Roussel et al.}{2003}]{roussel03}
    Roussel, H., Helou, G., Beck, R., Condon, J. J., Bosma, A.,
    Matthews, K., \& Jarrett, T. H.
    2003, ApJ, 593, 733
\bibitem[\protect\citeauthoryear{Rowlands et al.}{2014}]{rowlands14}
    Rowlands, K., et al.\ 2014, MNRAS, 441, 1017
\bibitem[Rubin et al.(1980)]{rubin80}	
	Rubin, V. C., Thonnard, N. \& Ford, W. K. Jr., 1980, ApJ, 238, 471
\bibitem[Rubin et al.(1985)]{rubin85}	
	Rubin, V. C. et al., 1985, ApJ, 289, 81
\bibitem[\protect\citeauthoryear{Sargent et al.}{1980}]{sargent80}
    Sargent, W. L. W., Young, P. J., Boksenberg, A., \& Tytler, D. 1980, ApJS, 42, 41
\bibitem[\protect\citeauthoryear{Saitoh et al.}{2009}]{saitoh09}
    Saitoh, T. R., Daisaka, H., Kokubo, E., Makino, J., Okamoto, T.,
    Tomisaka, K., Wada, K., \& Yoshida, N. 2009, PASJ, 61, 481
\bibitem[Sancisi(1983)]{sancisi83}	
	Sancisi, R., 1983, IAU symposium, 100, 55
\bibitem[Sancisi \& van Albada(1987a)]{sancisi1987a}	
	Sancisi, R. \& van Albada, T. S., 1987, IAU symposium, 117, 67
\bibitem[Sancisi \& van Albada(1987b)]{sancisi1987b}
	Sancisi, R. \& van Albada, T. S., 1987, "Dark matter" IAU symposium, 124, 699
\bibitem[Sandage et al.(1970)]{1970ApJ...160..831S}
	Sandage A. et al., 1970, ApJ, 160, 831
\bibitem[Scannapieco et al.(2005)]{scannapieco05}	
	Scannapieco E., Silk J. \& Bouwens R., 2005, ApJ, 635, 13
\bibitem[Schiminovich et al.(2007)]{schiminovich2007} 
	Schiminovich, D., Wyder, T.~K., Martin, D.~C., et al.\ 2007, ApJS, 173, 315 
\bibitem[Scoville et al.(2007)]{2007ApJS..172....1S} 
	Scoville, N., et al.\ 2007, ApJS, 172, 1 
\bibitem[di Serego et al.(2007)]{2007A&A...474..851D}
	di Serego Alighieri S. et al., 2007, A\&A, 474, 851
\bibitem[\protect\citeauthoryear{Seymour et al.}{2008}]{seymour08}
    Seymour, N., et al.\ 2008, MNRAS, 386, 1695
\bibitem[Serra et al.(2012)]{2012MNRAS.422.1835S}
	Serra P. et al., 2012, MNRAS, 422, 1835
\bibitem[Schiminovich et al.(2010)]{2010MNRAS.408..919S}
	Schiminovich D. et al., 2010, MNRAS, 408, 919
\bibitem[\protect\citeauthoryear{Smith, Cohen, \& Bradley}{Smith et al.}{1986}]{smith86}
    Smith, H. E., Cohen, R. D., \& Bradley, S. E. 1986, ApJ, 310, 583
\bibitem[\protect\citeauthoryear{Smol\v{c}i\'{c} et al.}{2014}]{smolcic14}
    Smol\v{c}i\'{c}, V., et al.\ 2014, MNRAS, 443, 2590
\bibitem[\protect\citeauthoryear{Smol\v{c}i\'{c} et al.}{2009}]{smolcic09}
    Smol\v{c}i\'{c}, V., et al.\ 2009, ApJ, 690, 610
\bibitem[Sofue et al.(1995)]{sofue95}	
	Sofue, Y., Honma, M. \& Arimoto, N., 1995, A\&A, 296, 33
\bibitem[Spitzer \& Baade(1951)]{1951ApJ...113..413S}
	Spitzer L. Jr. \& Baade W., 1951, ApJ, 113, 413
\bibitem[\protect\citeauthoryear{Srianand et al.}{2012}]{srianand12}
    Srianand, R., Gupta, N., Petitjean, P., Noterdaeme, P., Ledoux, C.,
    Salter, C. J., \& Saikia, D. J. 2012, MNRAS, 421, 651
\bibitem[Sternberg(1988)]{sternberg88}	
	Sternberg, A., 1988, ApJ, 332, 400
\bibitem[Struve \& Conway(2012)]{2012A&A...546A..22S}
    	Struve, C., \& Conway, J. E. 2012, A\&A, 546, 22
\bibitem[\protect\citeauthoryear{Symeonidis et al.}{2013}]{symeonidis13}
    Symeonidis, M., Vaccari, M., Berta, S., et al.\ 2013, MNRAS, 431, 2317
\bibitem[Takahashi et al.(1999)]{takahashi99}
	Takahashi, J., Masuda, K. \& Nagaoka, M., 1999, MNRAS, 306, 22
\bibitem[Takahashi(2000)]{takahashi2000}
 Takahashi, J.\ 2000, Astronomical Herald, 93, 191
\bibitem[\protect\citeauthoryear{Takeuchi et al.}{2001}]{takeuchi01}
    Takeuchi, T. T., Kawabe, R., Kohno, K., Nakanishi, K., Ishii, T. T.,
    Hirashita, H., \& Yoshikawa, K. 2001, PASP, 113, 783
\bibitem[Tanaka et al.(2014)]{tanaka14}
	Tanaka, A. et al., 2014, PASJ, 66, 66
\bibitem[Teruya \& Takeuchi(2014)]{teruya2014}
	Teruya, N. \& Takeuchi, T. T., 2014, in 
	``The Impact of Galactic Structure on Star Formation'', Sapporo, Japan
\bibitem[Thomas et al.(2010)]{2010MNRAS.404.1775T}
	Thomas D. et al., 2010, MNRAS, 404, 1775
\bibitem[Tinsley(1980)]{tinsley1980}
	Tinsley, B.~M.\ 1980, Fundam.\ Cosm.\ Phys., 5, 287 
\bibitem[Tinsley \& Danly(1980)]{tinsley_danly1980}
	Tinsley, B.~M., \& Danly, L.\ 1980, ApJ, 242, 435 
\bibitem[\protect\citeauthoryear{Totani \& Takeuchi}{2002}]{totani02}
    Totani, T., \& Takeuchi, T. T. 2002, ApJ, 570, 470
\bibitem[Tully \& Fisher(1977)]{tf77}	
	Tully, R. B. \& Fisher, J. R., 1977, A\&A, 54, 661
\bibitem[van den Bergh(1976)]{1976ApJ...206..883V}
	van den Bergh S., 1976, ApJ, 206, 883
\bibitem[van den Bosch \& Swaters(2001)]{2001MNRAS.325.1017V}
	van den Bosch F. C. \& Swaters R. A., 2001, MNRAS, 325, 1017
\bibitem[van Dishoeck \& Black(1986)]{vdb86}	
	van Dishoeck, E. F. \& Black, J. H., 1986, ApJS, 62, 109
\bibitem[van Gorkom \& Schiminovich(1997)]{1997ASPC..116..310V}
	van Gorkom J. \& Schiminovich D., 1997, ASPC, 116, 310
\bibitem[\protect\citeauthoryear{Vanzi et al.}{2008}]{vanzi08}
    Vanzi, L., Cresci, G., Telles, E., \& Melnick, J. 2008, A\&A, 486, 393
\bibitem[Verheijen et al.(2007)]{2007ApJ...668L...9V}
	Verheijen, M. et al., 2007, ApJ, 668, 9
\bibitem[\protect\citeauthoryear{Vladilo \& P\'{e}roux}{2005}]{vladilo05}
    Vladilo G., P\'{e}roux C. 2005, A\&A 444, 461
\bibitem[Voges et al.(1999)]{1999A&A...349..389V} 
	Voges, W. et al.\ 1999, ApJ, 756, L28 
\bibitem[\protect\citeauthoryear{Wada \& Norman}{2001}]{wada01}
    Wada K., Norman C. A., 2001, ApJ, 546, 172
\bibitem[Walter et al.(2008)]{walter08}	
	Walter, F. et al., 2008, AJ, 136, 2563
\bibitem[Wang et al.(2011)]{2011MNRAS.412.1081W}
	Wang J. et al., 2011, MNRAS, 412, 1081
\bibitem[Wardle \& Knapp(1986)]{1986AJ.....91...23W}
	Wardle M. \& Knapp G. R., 1986, AJ, 91, 23
\bibitem[Weldrake et al.(2003)]{2003MNRAS.340...12W}
	Weldrake D. T. F. et al., 2003, MNRAS, 340, 12
\bibitem[\protect\citeauthoryear{Wolfe \& Davis}{1979}]{wolfe79}
    Wolfe, A. M., \& Davis M. M. 1979, AJ, 84, 699
\bibitem[\protect\citeauthoryear{Wolfe et al.}{1985}]{wolfe85}
    Wolfe A. M., Briggs F. H., Turnshek D. A., Davis M. M., Smith H. E.,
    Cohen R. D. 1985, ApJ, 294, L67
\bibitem[\protect\citeauthoryear{Wolfe et al.}{1986}]{wolfe86}
    Wolfe, A. M., Turnshek, D. A., Smith, H. E., \& Cohen, R. D.
    1986, ApJS, 61, 249
\bibitem[Wolfe et al.(2013)]{wolfe13}	
	Wolfe, S. A. et al., 2013, Nature, 497, 224
\bibitem[Wong \& Blitz(2002)]{wb02}	
	Wong, T. \& Blitz, L., 2002, ApJ, 569, 157
 \bibitem[Wong et al.(2013)]{wong13}	
	 Wong, T. et al., 2013, ApJ, 777, 4
\bibitem[\protect\citeauthoryear{Yajima et al.}{2015}]{yajima15}
    Yajima, H., Shlosman, I., Romano-D\'{\i}az, E., \&
    Nagamine, K. 2015, MNRAS, 451, 418
\bibitem[York et al.(2000)]{2000AJ....120.1579Y}
	York D. G. et al., 2000, AJ, 120, 1579
\bibitem[Yoshida et al.(2006)]{2006ApJ...652....6Y}
    	Yoshida N. et al., 2006, ApJ, 652, 2
\bibitem[Yoshida et al.(2008)]{2008Sci...321..669Y}
   	Yoshida N. et al., 2008, Science, 321, 669
\bibitem[\protect\citeauthoryear{Yun, Reddy, \& Condon}{Yun et al.}{2001}]{yun01}
    Yun, M. S., Reddy, N. A., \& Condon, J. J. 2001, ApJ, 554, 803
\bibitem[Zaritsky et al.(2014)]{zaritsky14}	
	Zaritsky, D. et al., 2014, AJ, 147, 134

 
\end{thebibliography}

\end{document}